\documentclass[draft]{agujournal2019}
\usepackage{url}
\usepackage{amsmath}
\usepackage{lineno}
\usepackage{todonotes}
\usepackage[finalnew]{trackchanges} %for better track changes. finalnew option will compile document with changes incorporated. inline will show changes.
\usepackage{soul}

\usepackage{graphicx} % Required for inserting images
\usepackage{caption}
\usepackage{subcaption}

\journalname{Space Weather}

\begin{document}

%TC:ignore
\title{GIC-Related Observations During the May 2024 Geomagnetic Storm in the United States}
\authors{L. A. Wilkerson\affil{1}, R. S. Weigel\affil{1}, D. Thomas\affil{1}, D. Bor\affil{1}, E. J. Oughton\affil{1}, C. T. Gaunt\affil{2}, C. C. Balch\affil{3,4}, M. J. Wiltberger\affil{5}, and A. Pulkkinen\affil{6}}

\affiliation{1}{Space Weather Lab, George Mason University, Fairfax, VA}
\affiliation{2}{University of Cape Town, Cape Town, ZA}
\affiliation{3}{Cooperative Institute in Environmental Sciences (CIRES), Boulder, CO}
\affiliation{4}{NOAA/SWPC, Boulder, CO}
\affiliation{5}{NSF NCAR/HAO, Boulder, CO}
\affiliation{6}{NASA/GSFC, Greenbelt, MD}

\correspondingauthor{L. A. Wilkerson}{lwilker@gmu.edu}
%TC:endignore

%%%%%%%%%%%%%%%%%%%%%%%%%%%%%%%%%

%  List up to three key points (at least one is required)
%  Key Points summarize the main points and conclusions of the article
%  Each must be 140 characters or fewer with no special characters or punctuation and must be complete sentences
\begin{keypoints}
\item A large and unique set of GMD-related measurements are compared with model predictions and measurements
\item Empirical relationships between the maximum magnitude of GIC at a site and its location and local conductivity are considered
\item The maximum magnitude of GIC during this storm can be estimated using a linear relationship to NERC alpha and beta scaling factors

\end{keypoints}
\begin{abstract}

The May 2024 geomagnetic storm was one of the most severe in the past 20~years. Understanding how large geomagnetic disturbances (GMDs) impact geomagnetically induced currents (GICs) within electrical power grid networks is key to ensuring their resilience. We have assembled and synthesized a large and unique set of GMD-related data, compared model predictions with measurements, and identified empirical relationships for GICs in the contiguous United States for this storm. 
Measurement data include GIC data from $47$ sites and magnetometer data from $17$ sites.
Model data include GIC computed by the Tennessee Valley Authority (TVA) power system operators at $4$ sites, GIC computed using a reference model at $47$ sites, and the difference in the surface magnetic field from a baseline ($\Delta \mathbf{B}$) computed at $12$ magnetometer sites from three global magnetospheric models --- the Multiscale Atmosphere-Geospace Environment Model (MAGE), Space Weather Modeling Framework (SWMF), and Open Geospace General Circulation Model (OpenGGCM). 
GIC measured and computed by TVA had a correlation coefficient $\text{r}>0.8$ and a prediction efficiency between 0.4 and 0.7. The horizontal magnetic field perturbation from a baseline, $\Delta B_H$, computed by MAGE, SWMF, and OpenGGCM had a correlation r from $0.21$ to $0.65$. Two empirical relationships were considered: (1) how the correlation between measured GIC site pairs depended on differences in site separation distance, $\beta$ scaling factor (related to ground conductivity), and geomagnetic latitude; and (2) a regression model for the maximum $\mbox{GIC}$ magnitude at each site given the product of $\alpha$ (related to magnetic latitude) and $\beta$.
\end{abstract}

%TC:ignore
\section*{Plain Language Summary}

Geomagnetically induced currents (GICs) are a byproduct of geomagnetic storms, often caused by solar activity, and can impact critical Earth-based infrastructure, such as power grids. In May 2024, Earth experienced the largest geomagnetic storm in 20~years.
In this work, we examine a large and unique set of data during this event that includes measured and calculated values for both GIC and fluctuations in the magnetic field on Earth's surface. Measured GIC data are compared to GIC calculated by the Tennessee Valley Authority (TVA) and a reference model. Measured surface magnetic field data are compared to that calculated by three different global geospace simulation models. 
We also explore relationships between GIC measurement sites using a linear regression model. From this model, we find that the maximum magnitude of GIC during this storm can be estimated using a linear relationship with quantities related to geomagnetic latitude and ground conductivity.
%TC:endignore

\section{Introduction}

On 10--12 May 2024, NOAA Active Region 13664 triggered a geomagnetic storm (known as the ``Gannon" or ``Mother's Day" storm), which was the most intense geomagnetic storm in over two decades (\citeA{Kruparova24}; \citeA{TulasiRam2024}; \citeA{Kwak2024}).

Between 7--11 May 2024, NASA reported observing multiple strong solar flares, eight of which were X-class, and at least seven coronal mass ejections (CMEs) \cite{Johnson-Groh24}.
NOAA's Space Weather Prediction Center (SWPC) reported that the first CME encountered Earth on 10 May, resulting in a geomagnetic storm that reached a rating of G5 --- a level that had not occurred since October 2003 \cite{SWPC24_G5}. 
The storm had the largest Disturbance Storm Time index (\change[LW]{D$_{\mathrm{st}}$}{Dst}) magnitude and most negative \change[LW]{Sym-H}{SYM-H} since March 1989 (\citeA{SWPC24_Hist}; \citeA{TulasiRam2024}; \citeA{DeMichelis2025}) and reached the maximum K$_p$ index of 9 twice \cite{Zhang2025}.
On 11 May, the daily A$_\text{p}$-index was $273$ and the minimum \change[LW]{D$_{\text{st}}$}{Dst} was $-406$~nT, the largest single-day magnitudes recorded since December 2001 \cite{Mlynczak24}.
Solar wind conditions\add[LW]{ and select geomagnetic indices} during the storm\add[LW]{ --- including Dst, SYM-H, and K$_p$ ---}, retrieved from OMNI (\protect\citeA{OMNI2020_1hr}, \protect\citeA{OMNI2020_1min}) using the Heliophysics Application Programmer's Interface (HAPI; \protect\citeA{Weigel2021}), are illustrated in Figure~\protect\ref{fig:solar_wind}.
A statistical analysis by \citeA{Elvidge2025} indicates that the storm's magnitude, in terms of geomagnetic indices, was a 1-in-12.5-year event, while its duration was a 1-in-41-year event.

During the storm, aurora extended to middle-latitudes, with the southern aurora reaching the southern parts of Africa and South America \cite{Karan24}, and the northern aurora visible across much of the United States and Mexico (\citeA{SWPC24_G5}; \citeA{GonzalezEsparza2024}). The storm also caused interruptions to satellite operations --- particularly those in low Earth orbit (LEO) \cite{Parker24} --- and impacted GNSS/GPS services \cite{Themens2024}, for example, resulting in agricultural production losses in the Southwestern and Midwestern US (\citeA{Younas2025}; \citeA{Griffin2025}).

The main phase of the storm gave rise to large-amplitude geomagnetically induced current (GIC) fluctuations \cite{Waghule2025}.
Extreme ground geomagnetic field perturbations were produced between 01:50 UT and 02:30 UT in the American sector on 11 May \cite{Ngwira2025}.
In Mexico, GIC intensities of $\sim$30~A were observed at low-latitude (${\sim}25^\circ$--$40^\circ$ geomagnetic latitude) 400~kV substations, with effects lasting up to 2~h \cite{Caraballo25}; in New Zealand, 220~kV transformers experienced multiple short periods where GIC exceeded 50~A, maximizing at 113~A \cite{Clilverd2025}; and, in Russia, three strong GIC peaks of $\sim$50--62~A were observed in two 330~kV autotransformers \cite{Despirak2025}.

In this work, we present an analysis of the accuracy of two GIC models and three $\Delta\mathbf{B}$ models during the May 2024 storm. We also identify empirical relationships between GICs measured at different sites and the relationship between the maximum GIC magnitude during the storm and the $\alpha$ and $\beta$ scaling factors, related to geomagnetic latitude and ground conductivity, respectively (\citeA{NERC2020a}; \citeA{Pulkkinen25}; see also Section~\ref{sect:beta}).

In the remainder of this section, we review the current literature on space weather, GICs, and current methods and approaches to model validation.
In Section~2, we present a summary of the measurement and prediction data and results from the GIC and $\Delta\mathbf{B}$ models.
In Section~3, we present the results of the empirical analysis.

\subsection{GICs}
\label{sec:gic_intro}

GICs are caused by electric currents in Earth's ionosphere and magnetosphere. Enhancements in these currents are associated with geomagnetic storms and are often a direct result of solar activity, such as CMEs (\citeA{Winter19}; \citeA{Gopalswamy2022}). These eruptive solar events send outflows of charged particles through interplanetary space, which can transfer energy to Earth's magnetospheric-ionospheric system, resulting in current enhancements that cause geoelectric field variations on Earth and, as a result, GICs (\citeA{Pulkkinen17}; \citeA{DeMichelis2025}). Extreme events, such as the Carrington Event (\citeA{Carrington1859}; \citeA{Blake2020}; \citeA{Blake2021a}; \citeA{Thomas2024}), have the potential to produce geoelectric field variations more than twice as large as those produced by previously observed Earth-directed extreme events, such as the March 1989 storm \cite{Ngwira14}.

The strength of the geoelectric field, which is the driver of GICs, at a particular point on Earth's surface depends on geomagnetic latitude and ground conductivity \cite{Wawrzaszek2024}. Independent of ground conductivity, higher geomagnetic latitude locations are more likely to experience stronger GICs because they are nearer to the auroral-zone current systems, which tend to be larger and more variable (\citeA{Winter19}; \citeA{Hajra2022}). However, significant GICs can occur at low- to middle-latitudes due to their dependence on ground conductivity, especially during intense geomagnetic events when the auroral zone expands (\citeA{Liu18}; \citeA{Caraballo23}).
Global magnetohydrodynamics (MHD) models of extreme geomagnetic events predict a strongly shifted auroral geomagnetic latitude boundary, implying that the region of large induced ground electric fields may be displaced towards the equator, affecting power grids in regions normally far away from the auroral zone, such as the southern United States and central/southern Europe \cite{Blake2021b}.
%Additionally, locations with less surface conductivity experience stronger GICs as currents flow more freely through resistive material (\cite{Wawrzaszek2024}).

Large geomagnetic storms can negatively impact terrestrial critical infrastructure, for example, causing radio blackouts \cite{Knipp16} and phone system outages (\citeA{Boteler98}; \citeA{Love2025}). 
GICs, in particular, are known to impact critical infrastructure that relies on long systems of electrical conductors, such as pipelines (\citeA{Khanal19}; \citeA{Ingham2025}), railway signaling and switching systems (\citeA{Love2019}; \citeA{Boteler2020}; \citeA{Patterson2024}), submarine cables \cite{Boteler24}, and electrical power transmission networks (\citeA{Kappenman96}; \citeA{MacManus2025}). 
The interaction between power system infrastructure and GIC can result in damaged equipment, such as transformers (\citeA{Gaunt07}; \citeA{Gaunt2016}), cascading failures leading to voltage collapse \cite{Xin2025}, and even system-wide disturbances, as seen in the 1989 Hydro-Québec blackout \cite{Bolduc02}. 
Extra high-voltage (EHV) transformers in particular are a concern when it comes to the potential catastrophic damage of a geomagnetic storm, as there are $\sim$2,000~EHV transformers in the United States (\citeA{MacAlester14}; \citeA{Lamichhane2022}).

Furthermore, the potential economic impact due to power grid failures in the US has been estimated to be a daily loss of GDP ranging from $6.2$ to $42$ billion US dollars \cite{Oughton17}, $2.9$ to $15.9$ billion pounds in the UK \cite{Oughton2018}, and $3$ to $8.36$ NZ dollars in New Zealand \cite{Oughton2025a}. Given the potential for far-reaching impact on vital infrastructure and the global economic systems, both accurate prediction and preparation for extreme GIC events are essential. As such, power systems operators require modeling and prediction capabilities to inform their decision-making during these events \cite{Aslani2024}.

%However, empirical GIC estimates are sensitive to network configuration during the event, and this perishable network data is often lost after a GMD. 

\subsection{GIC Modeling Overview}

The strength of GICs and their impact on infrastructure at any given location on Earth's surface depend on several variables, namely latitude, ground conductivity, and grid configuration \cite{Lanabere23}. 
Current methods of GIC calculations involve transmission lines at different AC voltage levels, resistances of individual transformers, characteristics of the geomagnetic source fields, and 3-D Earth conductivity structure (\citeA{Boteler2017}; \citeA{Rosenqvist2025}).
%In the US, the ground conductivity structure results in more than three orders of magnitude difference in geoelectric hazards \cite{Lucas20}.
Layered 1-D piecewise conductivity models have historically been demonstrated to be a useful and relatively simple method for describing approximate ground conductivity (e.g., \citeA{Gannon17}, \citeA{Caraballo23}).
%These 1-D models, along with latitude and geomagnetic field data, are then used to find the geoelectric field, which is used to compute GIC at geographical points. 
However, the accuracy of these 1-D models is limited (\citeA{Weigel2017}; \citeA{Caraballo23}; \citeA{Balch23}). 
3-D conductivity models, in contrast, have been found to better account for deviations in GIC, as they provide a more accurate representation of geologic complexity (\citeA{Kelbert20}; \citeA{AlvesRibeiro23}, \citeA{Balch24}).
%A comparison study of ground electric field calculations using both 1-D and 3-D conductivity models found that, while 3-D conductivity models are more accurate in terms of representing the spatial variation of the peak geoelectric field across a given region, 1-D approximations can still offer an accurate average induction response over the region \cite{Gannon17}.

Predicting the geomagnetic field is required for predicting GIC. However, current operational global MHD and geospace environment models do not fully capture the ground-level magnetic field variability, which is essential for modeling induction hazards (\citeA{Haines22}; \citeA{Florczak2023}).

While GIC can impact various essential infrastructure systems, its impact on power grid networks is of particular interest, especially after the Hydro-Quebec outage in March of 1989, which led to a 9~h blackout and resulted in $\sim13.2$ million Canadian dollars in total power grid network damages (\protect\citeA{Allen1989}; \citeA{Bolduc02}; \citeA{Hughes2022}). 
As such, the work of estimating GICs naturally lends itself to estimating their impact on power grid networks. 
In addition to ground conductivity and geomagnetic location, variables such as power line network orientation (north-south vs. east-west), operational voltage, and line length can impact the extent of damage caused by GICs \cite{Caraballo23}.
%\todo[]{Do they have results we can compare with?}.
To accurately estimate substation-level GICs for a given power grid network, the first step is to find the induced voltages by integrating the surface geoelectric fields along power lines \cite{Kelbert20}. Then, knowledge of the line resistance, the line connectivity to substations, and the current flow through the substations is needed to calculate GIC in the transformers. These calculations can then be compared with a measurement to perform a model validation assessment \cite{Shetye18a}. 

%This paper will consider GICs in the contiguous US and their effects on the US power grid, particularly in the Tennessee Valley region.

\section{Summary of Data}
\label{sect:summary}

\subsection{Measurements}

\begin{figure}[h]
    \caption{Solar wind conditions and select geomagnetic indices during the May 2024 storm event from the OMNI dataset (\protect\citeA{OMNI2020_1hr}, \protect\citeA{OMNI2020_1min}) using their HAPI \protect\cite{Weigel2021} server. AE is the Auroral Electrojet index, and AL is the Auroral Electrojet index, Lower. The K$_p$ index is based on the surface $\mathbf{B}$ field variation over 3~h intervals, with values ranging from 0--9. The SYM-H index, related to Dst, is a measure of the deviation of the horizontal (H) component of the $\mathbf{B}$ field. T is the temperature of the proton component of the thermal solar wind. Mag Mach (magnetosonic Mach number) is a dimensionless quantity representing the ratio of the solar wind velocity to the local Alfvén speed. V$_x$ is the $x$ component of the solar wind velocity. B$^{\text{IMF}}_y$ and B$^{\text{IMF}}_z$ are the $\mathbf{B}$ field components of the IMF in the $y$ and $z$ directions, respectively.}
    \centering
    \includegraphics[width=.9\textwidth]{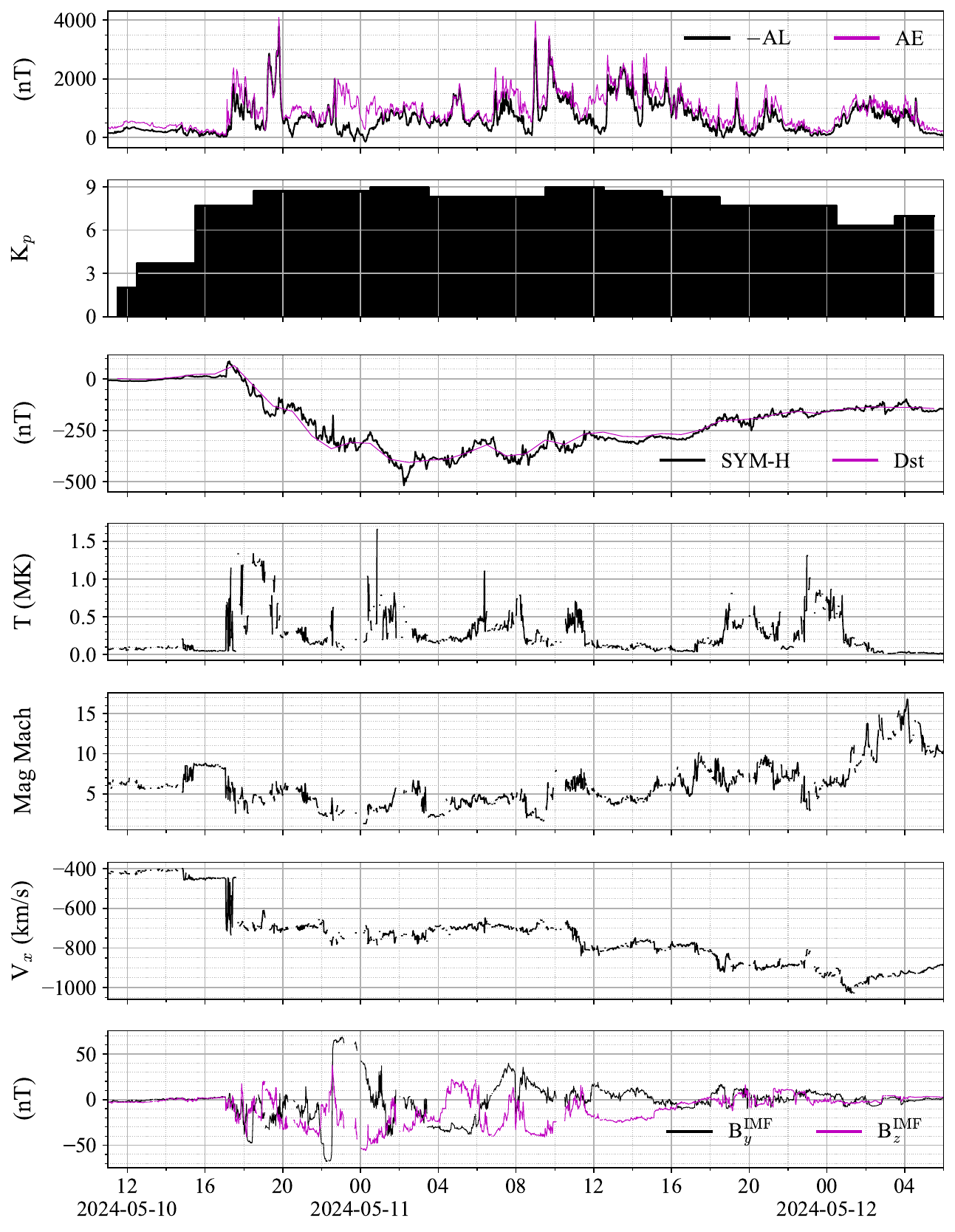}
    \label{fig:solar_wind}
    %\note[LW]{Figure 1 updated to use data for full storm OMNI data retrieved using HAPI. Also added Dst to SYM-H plot.}
\end{figure}

In Figure~\ref{fig:solar_wind}, the near-Earth solar wind measurements and geomagnetic indices for the event, used as inputs for the geospace environment models described in Section~\protect\ref{sect:delta-b}, are shown.
The modeled and measured GIC and ground magnetic field ($\mathbf{B}$) timeseries used in this work at locations throughout the contiguous United States during the May 2024 storm are summarized in Figure~\ref{fig:loc_map}. The Tennessee Valley Authority (TVA)-provided measured and modeled GIC and measured $\mathbf{B}$ were obtained directly from them \cite{Balch24}; the North American Electric Reliability Corporation (NERC) data were obtained from the NERC Electric Reliability Organization (ERO) data portal \cite{NERC24}; and the simulation data from the Space Weather Modeling Framework (SWMF), Multiscale Atmosphere-Geospace Environment Model (MAGE), and Open Geospace General Circulation Model (OpenGGCM or GGCM) are described in Section~\ref{sect:delta-b}.

Each measured GIC timeseries was visually inspected for significant errors. Observed errors included a non-zero baseline, large and unphysical spikes, and time intervals of constant GIC values lasting longer than $2$~min. Additionally, any GIC timeseries with a cadence $\ge1$~min or a gap in the recorded data longer than $5$~min were omitted. 
Of the $17$~TVA and $396$~NERC GIC sites, $2$~TVA and $355$~NERC datasets were removed based on visual inspection. Also, $9$~NERC timeseries were duplicates of those provided by TVA. As a result, in this work, we consider data from $47$ GIC measurement sites.
In Figure~\ref{fig:gic_tva}, the measured GIC for the $15$~TVA sites is shown.
In Figure~\ref{fig:gic_nerc}, GIC data from NERC's ERO portal not found in TVA's data set are shown. In Figure~\protect\ref{fig:db_all}, measured $\Delta B_H$ data from both TVA and NERC magnetometers is shown.
%$15+(396-355-9)=15+32=47$

\add[LW]{Examining the measured solar wind and geospace parameters in Figure~\protect\ref{fig:solar_wind}, sudden storm commencement occurs with the sudden increase of SYM-H (and Dst) at 2024-05-10 17:15. 
Around this time, the Auroral Electrojet index (AE) and the negative Auroral Electrojet index, Lower, (-AL) both rapidly increase; the K$_p$ index jumps from 3.7 to 7.7; the temperature of the proton component of the thermal solar wind (T) sharply increases and fluctuates rapidly; the $x$ component of the
solar wind velocity (V$_x$) sharply decreases and fluctuates rapidly; and the $y$ and $z$ components of the IMF $\mathbf{B}$ field (B$^{\text{IMF}}_y$ and B$^{\text{IMF}}_z$, respectively) are more variable.
This sudden storm commencement aligns with GIC measurements deviating from their baseline values at all TVA sites (Figure~\protect\ref{fig:gic_tva}) and most NERC sites (Figure~\protect\ref{fig:gic_nerc}). 
Similarly, in Figure~\protect\ref{fig:db_all}, the magnitude of $\Delta B_H$ begins to increase around this time at all sites.
Additionally, the minimum Dst value, occurring at 2024-05-11 02:30, and the minimum SYM-H value, occurring at 2024-05-11 02:14, are concurrent with a brief period of increased GIC activity (Figures~\protect\ref{fig:gic_tva} and \protect\ref{fig:gic_nerc}), with this concurrence being more prominent for GIC monitor sites that generally observed less GIC activity over the full storm duration. Likewise, a peak in $\Delta B_H$ is observed at this time at a majority of magnetometer sites (Figure~\protect\ref{fig:db_all}), with the peak magnitude increasing with magnetic latitude.}

\subsection{Transmission Lines}
\label{sect:transmission-lines}

Transmission line geographic and voltage information from the U.S. Department of Homeland Security's Homeland Infrastructure Foundation-Level Data (HIFLD) was used to estimate the power line AC voltage on which the GIC monitors were mounted (\citeA{hifld2023}; see Section~\ref{sect:intra}). The shapefile contains information on transmission line geography in the contiguous US, along with line voltage levels.
%https://hifld-geoplatform.hub.arcgis.com/datasets/3486fb60feb2454c99232248fdf624ec_0/explore?location=37.834593%2C-93.585144%2C4.78
% Consider adding NERC regions?
%https://hifld-geoplatform.hub.arcgis.com/maps/6b2af23c67f04f4cb01d88c61aaf558a/about

\subsection{GIC Estimation}
\label{sect:gic-estimation}

\subsubsection{TVA Calculation}
\label{sec:tva-calc}

One~min cadence $\mathbf{B}(t)$ measurements from nine U.S. Geological Survey (USGS; \citeA{USGS1985}) magnetometers (BOU, BSL, CMO, FRD, FRN, NEW, SIT, SJG, TUC), nine Natural Resources Canada (NRCan; \citeA{Newitt2007}) magnetometers (BLC, BRD, FCC, MEA, OTT, SNK, STJ, VIC, YKC), and six TVA magnetometers (ACK, BLR, PAR, RCM, UNI, WTB) were used to compute the parameters of a spherical elementary current system (SECS) model (\citeA{Amm1998}; \citeA{Pulkkinen2014}; \citeA{Vanhamki2019}). The SECS model currents were then used to compute $\Delta\mathbf{B}(t)$ at the locations where magnetotelluric transfer (MT) functions are available from the U.S. MT survey sites in the USArray/IRIS data portal \cite{Schultz2018}. At these sites, the MT transfer functions (\citeA{Kelbert2011}; \citeA{Kelbert2018}) were convolved with the SECS $\Delta\mathbf{B}(t)$ to estimate $\mathbf{E}(t)$ at the MT sites. The $\mathbf{E}(t)$ calculation is done in the time domain using a $1024$~min window and a causal convolution technique \cite{Boteler2022}. $\mathbf{E}(t)$ was then resampled with bilinear interpolation to a regular grid with a $0.5^\circ$ resolution in latitude and longitude, and this timeseries was used as input by TVA analysts to compute GIC flows based on their representation of the power grid.

%\todo[inline]{Chris: Is the list of magnetometers used the same as used by the Reference model?} 
%\todo[inline]{Chris: Need date as requested by ds.iris.edu/ds/products/emtf/}

\subsubsection{Reference Model Calculation}

The Reference model calculation presented in this section is a GIC model that can be used when no detailed information about the power network system is available to researchers (i.e., lines in service, line resistances, and substation circuit particulars; additional implementation details and results are given in \citeA{Oughton2025b}).
Because the Reference model, unlike the TVA model, does not contain information on the real-time power system configuration, it is expected to have less accuracy.

In this calculation, 1~min cadence $\mathbf{B}(t)$ measurements from 10 USGS and 14 NRCan magnetometers were used to compute the parameters of a SECS model, as in \citeA{Lucas20}. The magnetometer site IAGA codes are ALE, BOU, BLC, BRD, BRW, BSL, CBB, CMO, DED, FCC, FRD, FRN, IQA, MEA, NEW, OTT, RES, SHU, SIT, SJG, STJ, TUC, VIC, and YKC. (Note that this list and the resulting calculations differ from those in Section~\ref{sec:tva-calc} because independent groups provided the results.) 
Magnetometer data were visually checked for baseline offsets and spikes (none found), and the baseline was removed by subtracting the median value of each component during a two-day time interval starting 12:00 UTC on 10 May.
Similar to the TVA calculations in Section~\ref{sec:tva-calc}, the SECS model currents are then used to compute $\Delta\mathbf{B}(t)$ at the locations where magnetotelluric transfer (MT) functions are available from the U.S. MT survey sites in the USArray/IRIS data portal (retrieved in May 2024). At the MT sites, $\mathbf{E}(f)$ was derived by convolving $\Delta\mathbf{B}(f)$ with the transfer functions in the frequency domain. The frequency-domain geoelectric field, $\mathbf{E}(f)$, was then transformed into the time domain using the discrete Fourier transform to obtain $\mathbf{E}(t)$ at each site. The induced voltages were computed using \texttt{bezpy} Python routines \cite{Lucas2023} with inputs of $\mathbf{E}(t)$ and the transmission lines described in Section~\ref{sect:transmission-lines} and the Delaunay interpolation option for interpolating the electric field onto the transmission lines (as in \protect\citeA{Bonner2017}, \citeA{Kelbert20}).
This interpolation method may fail to capture sharp conductivity boundaries and local-scale conductivity variations.

The power line geometry and voltages used are described in Section~\ref{sect:transmission-lines}. As in \protect\citeA{Oughton2025b}, parameter values, including transformer winding resistances, transmission line resistances, and substation grounding, from the \citeA{Horton2012} benchmark model were used to construct a system admittance matrix.
To solve for the resulting GICs in the power network, the method by \protect\citeA{Pirjola2022} was used.
%\add[LW]{The controlled Horton test confirmed that our methods reproduce published results.}

\subsubsection{Results}

\begin{figure}[h]
     \centering
     \begin{subfigure}[b]{0.49\textwidth}
         \centering
         \includegraphics[width=\textwidth]{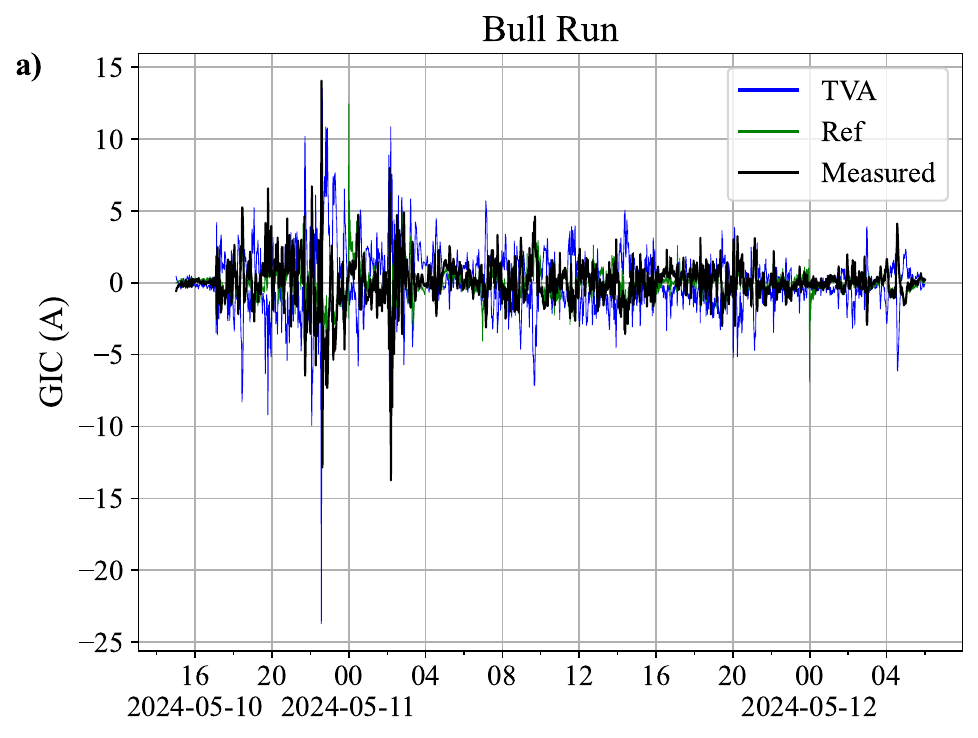}
         \label{br_gic}
     \end{subfigure}
     \begin{subfigure}[b]{0.49\textwidth}
         \centering
         \includegraphics[width=\textwidth]{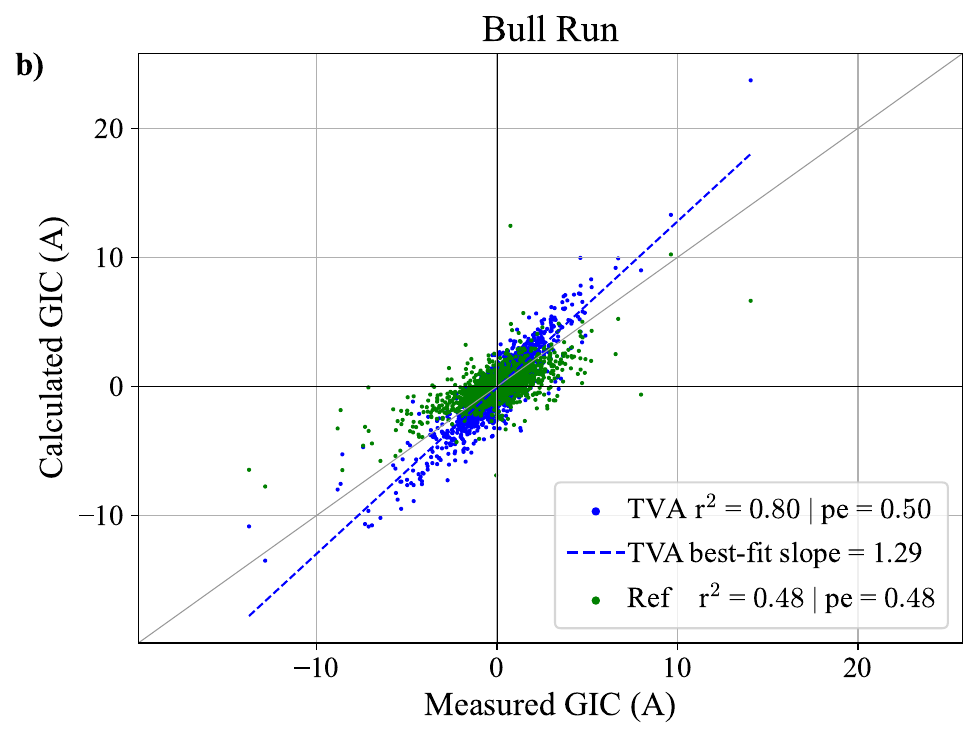}
         \label{br_cor}
     \end{subfigure}
     \begin{subfigure}[b]{0.49\textwidth}
         \centering
         \includegraphics[width=\textwidth]{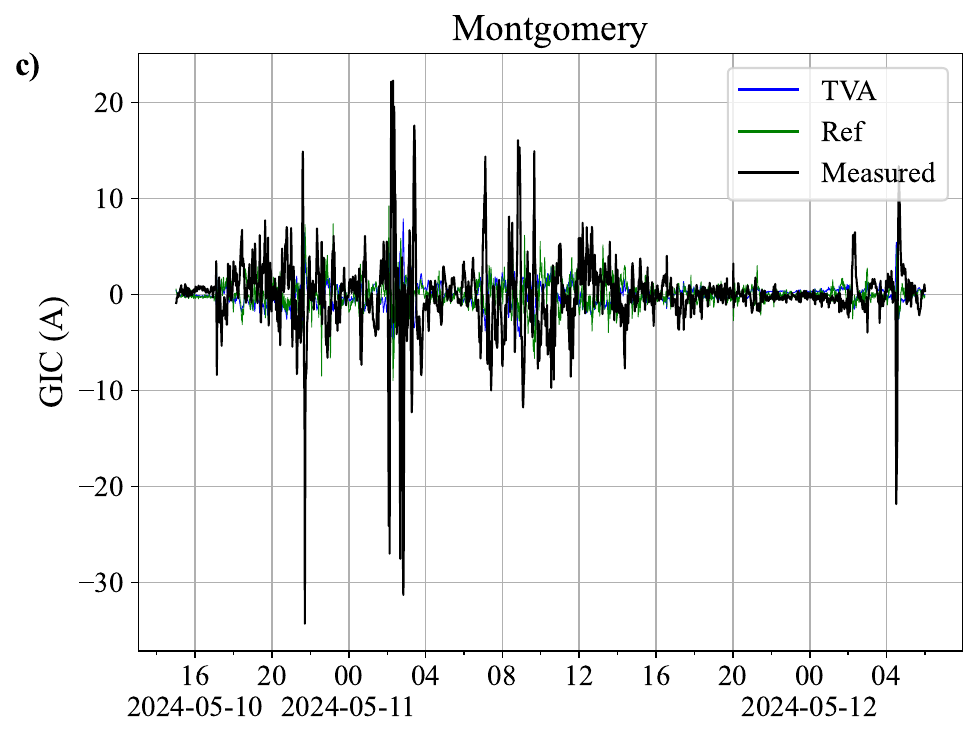}
         \label{mg_gic}
     \end{subfigure}
     \begin{subfigure}[b]{0.49\textwidth}
         \centering
         \includegraphics[width=\textwidth]{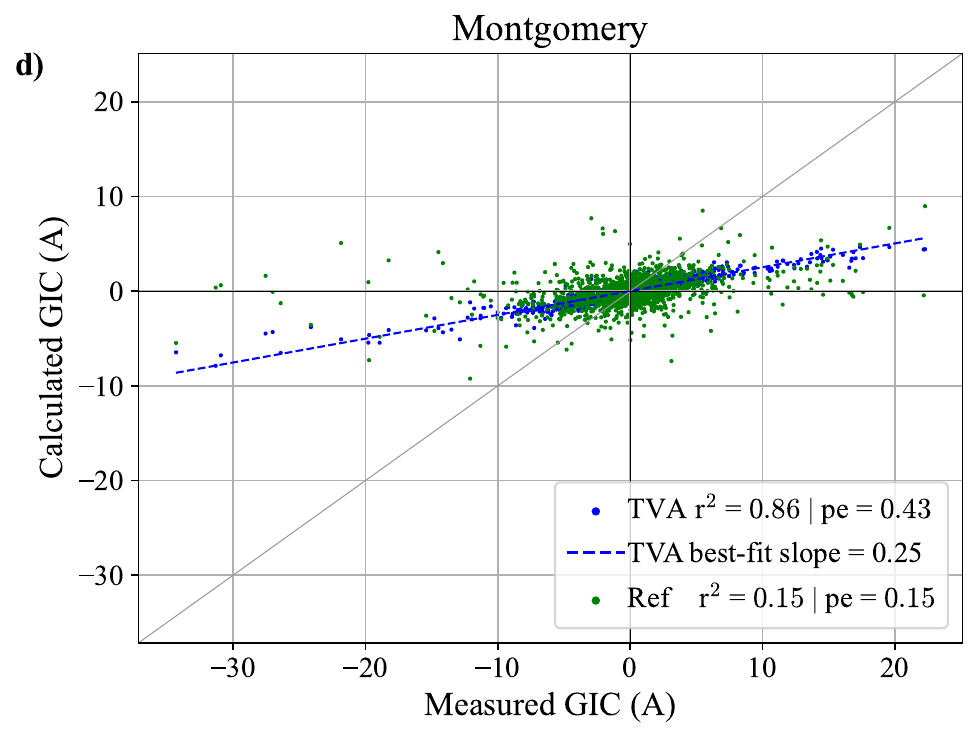}
         \label{mg_cor}
     \end{subfigure}
     \begin{subfigure}[b]{0.49\textwidth}
         \centering
         \includegraphics[width=\textwidth]{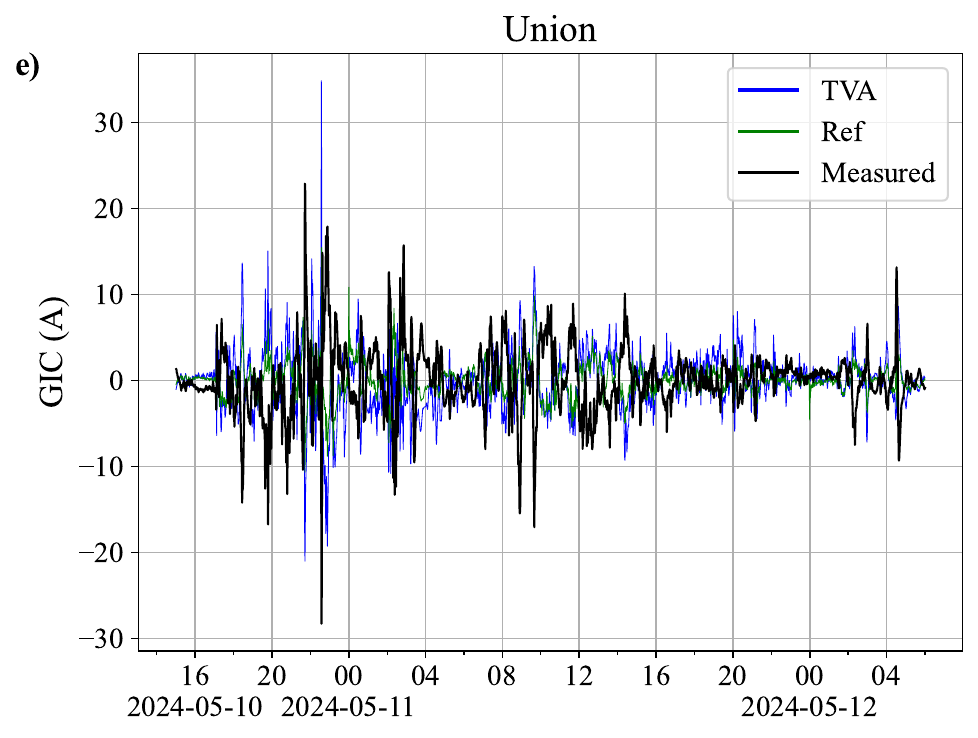}
         \label{un_gic}
     \end{subfigure}
     \begin{subfigure}[b]{0.49\textwidth}
         \centering
         \includegraphics[width=\textwidth]{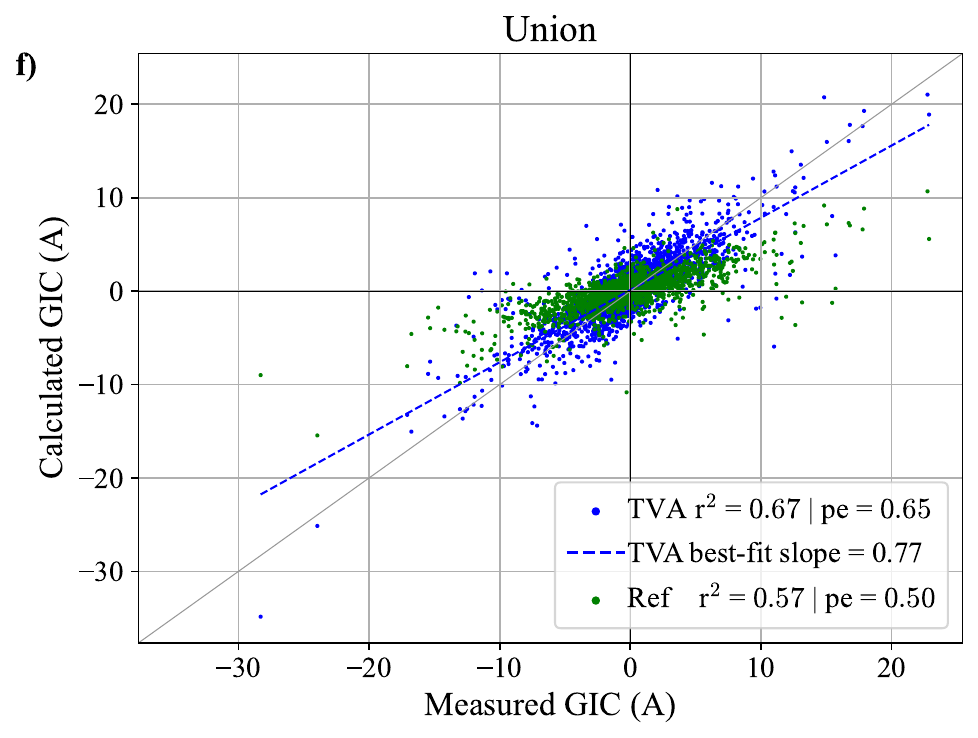}
         \label{un_cor}
     \end{subfigure}
     \begin{subfigure}[b]{0.49\textwidth}
         \centering
         \includegraphics[width=\textwidth]{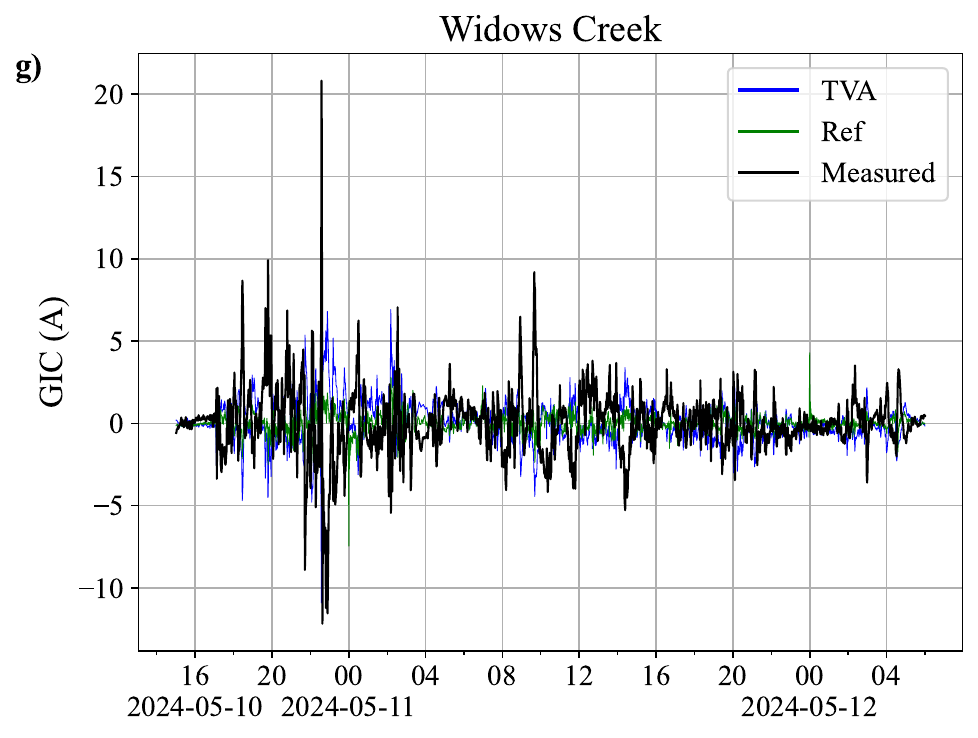}
         \label{wc_gic}
     \end{subfigure}
     \begin{subfigure}[b]{0.49\textwidth}
         \centering
         \includegraphics[width=\textwidth]{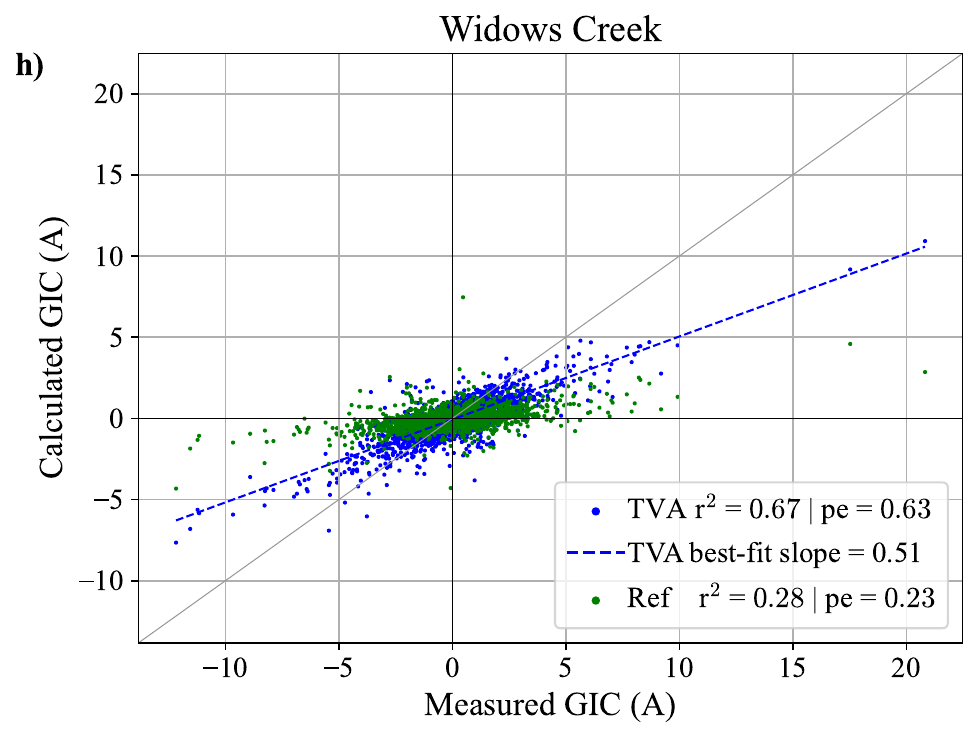}
         \label{wc_cor}
     \end{subfigure}
        \caption{Measured GIC, TVA model, and Reference model timeseries (left) and scatter correlation plots (right)}.
        \label{fig:gmu}
    %\note[LW]{Updated Fig 6 timeseries plots (left) to have "Measured" on top. All plots have been updated with results from most recent Reference model run.}
\end{figure}

In Figure~\ref{fig:gmu}, we compare the measured GIC with the TVA and Reference model calculations at the four sites where TVA modeled GIC was also available (Bull Run, Montgomery, Union, and Widows Creek; see Figure~\ref{fig:loc_map}). The TVA model predictions are at a 1~min cadence, so the 1~s measured GIC was averaged using 1~min non-overlapping windows. 

Three metrics were used to assess predictions: the Pearson correlation coefficient, $\mbox{r}$, the coefficient of determination, $\mbox{r}^2$, and the prediction efficiency \cite{Liemohn2018}, defined as $\mbox{pe}=1-\sum_t(\mbox{GIC}_{m}(t)-\mbox{GIC}_{p}(t))^2/\sum_t(\mbox{GIC}_{m}(t)-\overline{\mbox{GIC}}_{m})^2$, where subscript $m$ is for measured and subscript $p$ is for predicted (i.e., modeled) and $t$ is the index of the 1~min interval from 2024-05-10T15Z to 2024-05-12T06Z. 
The prediction efficiency is a skill metric relative to sample climatology (a prediction that is the mean of the measured) and depends on the error variance of the modeled timeseries divided by the variance of the measured timeseries. $\mbox{pe}=1$ indicates that the model exactly predicts the observations (error variance is 0); $\mbox{pe}=0$ indicates the model predicts as well as the mean of the observations. One advantage of $\mbox{pe}$ over $\mbox{r}$ (and $\mbox{r}^2$) is that $\mbox{pe}$ is sensitive to errors in magnitude \cite{Detman1997}. When the means and variances of the measured and predicted data are equal, the prediction error is uncorrelated with the predictor, and the mean of the prediction error is zero, $\mbox{pe}=\mbox{r}^2$. As such, our analysis will utilize all three of these metrics: $\mbox{r}$ as a measure of correlation vs. anti-correlation as defined by covariance, $\mbox{r}^2$ as a measure of goodness of fit, and $\mbox{pe}$ as a measure of predictive skill.

A summary of the prediction results is shown in Table~\ref{table:four_sites} for the four sites shown in Figure~\ref{fig:gmu}. At these four sites, both models predict GIC with a non-zero correlation. As expected, the TVA model outperforms the Reference model. 

Although the TVA model produces $\mbox{r}^2 > 0.66$ ($\mbox{r} > 0.81$) at all sites, the scatter plots in the right column of Figure~\ref{fig:gmu} show that the model over-predicts $|$GIC$|$ at Bull Run and under-predicts at Montgomery and Widows Creek.

\begin{table}[h]
\centering
\begin{tabular}{lrrrrrrr}
Site ID & $\sigma$(A) & $\sigma_\text{TVA}$ & $\sigma_\text{Ref}$ & $\text{r}^2_\text{TVA}$ & $\text{r}^2_\text{Ref}$ & $\text{pe}_\text{TVA}$ & $\text{pe}_\text{Ref}$ \\
\hline
Bull Run & 1.5 & 2.2 & 1.1 & $0.80\phantom{*}$ & $0.48\phantom{*}$ & $0.50*$ & $0.48\phantom{*}$ \\
Montgomery & 3.7 & 1.0 & 1.3 & $0.86\phantom{*}$ & $0.15\phantom{*}$ & $0.43*$ & $0.15*$ \\
Union & 3.8 & 3.5 & 1.8 & $0.67\phantom{*}$ & $0.57\phantom{*}$ & $0.65*$ & $0.50*$ \\
Widows Creek & 1.9 & 1.2 & 0.6 & $0.67\phantom{*}$ & $0.28\phantom{*}$ & $0.63*$ & $0.23*$ \\
\end{tabular}
\caption{Prediction metrics for TVA and Reference model. When r is negative, the values of pe are computed by multiplying the measured GIC by $-1$, which assumes the GIC sensors are installed in an orientation opposite to that assumed by the models \protect\cite{Balch24}. These cases are indicated 
with a *.}
\label{table:four_sites}
%\note[LW]{Table has been updated with metrics from most recent Reference model run.}
\end{table}

The Reference model GIC was also calculated at the $39$ other GIC measurement sites. Across the total of $43$ GIC sites, the Reference model GIC had a correlation coefficient ranging from $0.01$ to $0.85$ and a prediction efficiency ranging from $-36.77$ to $0.59$. Prediction metrics at all sites where GIC measurements are available are shown in Table~\ref{table:all_sites_GIC} of the Appendix.

The results in this work regarding the accuracy of GIC models are comparable to those found in recent literature, as summarized below. Although the approach, events, and metrics for model accuracy differ, these results provide context for the metrics found in this work.

% ccs in https://colab.research.google.com/drive/1alnSiEAOVhTak_vsqcronR87xeg7gze2?usp=sharing#scrollTo=bBK4AbFCCZ2G

For the May 2024 storm, \citeA{Caraballo25} computed GIC at three sites in Mexico using a conductivity and power grid model and found $\mbox{r}^2$ that ranged from $0.01$ to $0.33$. \citeA{Shetye18a} computed GIC for another storm (June 2015) at five sites with GIC instruments (with locations withheld) in the TVA power system network using ground magnetic field measurements at Fredericksburg, which is $\sim$600~miles (965~km) from the center of the TVA network. The comparison was made over a 2~h period in June 2015. They used both 1-D and 3-D ground conductivity models to calculate the electric field based on this magnetic field and applied the resulting electric field to two different power system network models. Although they did not provide summary metrics for the model predictions, visually, the model predictions shown in Figure~\ref{fig:gmu} appear to be comparable to those in \citeA{Shetye18a}.
\protect\citeA{Butala2017} compared measured GIC provided by the American Transmission Company (ATC) with two GIC models, one from 1-D transfer functions and the other from 3-D, for three GMD intervals in 2013. While no summary metrics were explicitly provided, the differences between measured and modeled timeseries peaked at $\sim$6~A, and both models failed to capture enhancements.
\citeA{Shetye18b} compared modeled and measured GIC using measured GIC from two power systems, one in the Eastern US and another in the Western US. 1-D and 3-D ground conductivity models were used, and the $\mbox{r}^2$ found ranged from $0.07$ to $0.65$. \citeA{Weigel2019} evaluated several models that used direct measurements of the geoelectric field in Japan to predict GIC and found pes in the range of $0.60$ to $0.83$. \citeA{Balch24} computed GIC at 12 sites in the TVA network for a 23.5~h period starting on 23 March 2023, at 00:30 UT, and found $\mbox{r}^2$ values ranging from 0.21--0.96 using the NOAA/USGS 3-D Geoelectric Field and a network simulation.

\subsection{$\Delta \mathbf{B}$ Models}
\label{sect:delta-b}

We use three geospace environment models to simulate the magnetic field measured on Earth. The models are MAGE (\citeA{sorathia2023multiscale}; \citeA{Merkin2025mage}), SWMF (\citeA{Toth2005}; \citeA{Gombosi2021swmf}), and OpenGGCM \cite{Raeder2008openggcm}. As global MHD models, they are broadly similar. They model the magnetosphere, field-aligned currents, and the ionosphere. However, the implementation details vary substantially between the three. They use different coordinate systems, MHD equations (e.g, symmetric versus semi-conservative), numerical differential equation solvers, models of the inner magnetosphere, etc. Thus, we do not expect the three models to produce identical simulation results.

The SWMF and OpenGGCM runs were executed at the Community Coordinated Modeling Center (CCMC; \citeA{Hesse2001}) using raw solar wind conditions from the Deep Space Climate Observatory (DSCOVR; \citeA{Burt2012}; \citeA{dscovr2016}). 
\change[LW]{The MAGE runs were executed at the Center for Geospace Storms (CGS) using DSCOVR—based solar wind conditions.}{The
% As described by Pham et al. 2026
MAGE runs were executed by the Center for Geospace Storms and used solar wind inputs constructed from the Magnetospheric Multiscale (MMS; \protect\citeA{Burch2015}) Mission and Acceleration, Reconnection, Turbulence and Electrodynamics of the Moon's Interaction with the Sun (ARTEMIS; \protect\citeA{Angelopoulos2010}) spacecraft observations, which were located much closer the magnetosphere than the L1 position providing a more accurate representation of the solar wind impacting geospace.}
\add[LW]{During the time interval of 2024-05-10 12:00 through 2024-05-11 12:00, the RMSE difference between the input V$_x$, B$_y^\text{IMF}$, and B$_z^\text{IMF}$ between those used by MAGE vs SWMF/OpenGGCM was 59.8 km/s, 7.32 nT, and 7.54 nT, respectively. (A comparison of these three parameters is in Figure~\protect\ref{fig:compare_solar_wind} in the Appendix.)}

\begin{figure} [h]
    \caption{GIC and magnetometer (measured difference in the surface magnetic field from a baseline, or $\Delta \mathbf{B}$) locations and locations where model estimates are available. The approximate TVA operating region is shown in the yellow box.The inset table describes the measurement type, measurement source, cadence ($\Delta\mbox{t}$), and number of sites (\#) for each symbol. *Note that $\Delta\mbox{t}$ for NERC GIC and $\Delta\mathbf{B}$ measured sites varied, with GIC sites having cadences of 1~s, 2~s, 4~s, 5~s, 1~min, 5~min, and 1~h and $\Delta\mathbf{B}$ sites having cadences of 1~s, 10~s, and 1~min.}
    \centering
    \includegraphics[width=1.0\textwidth]{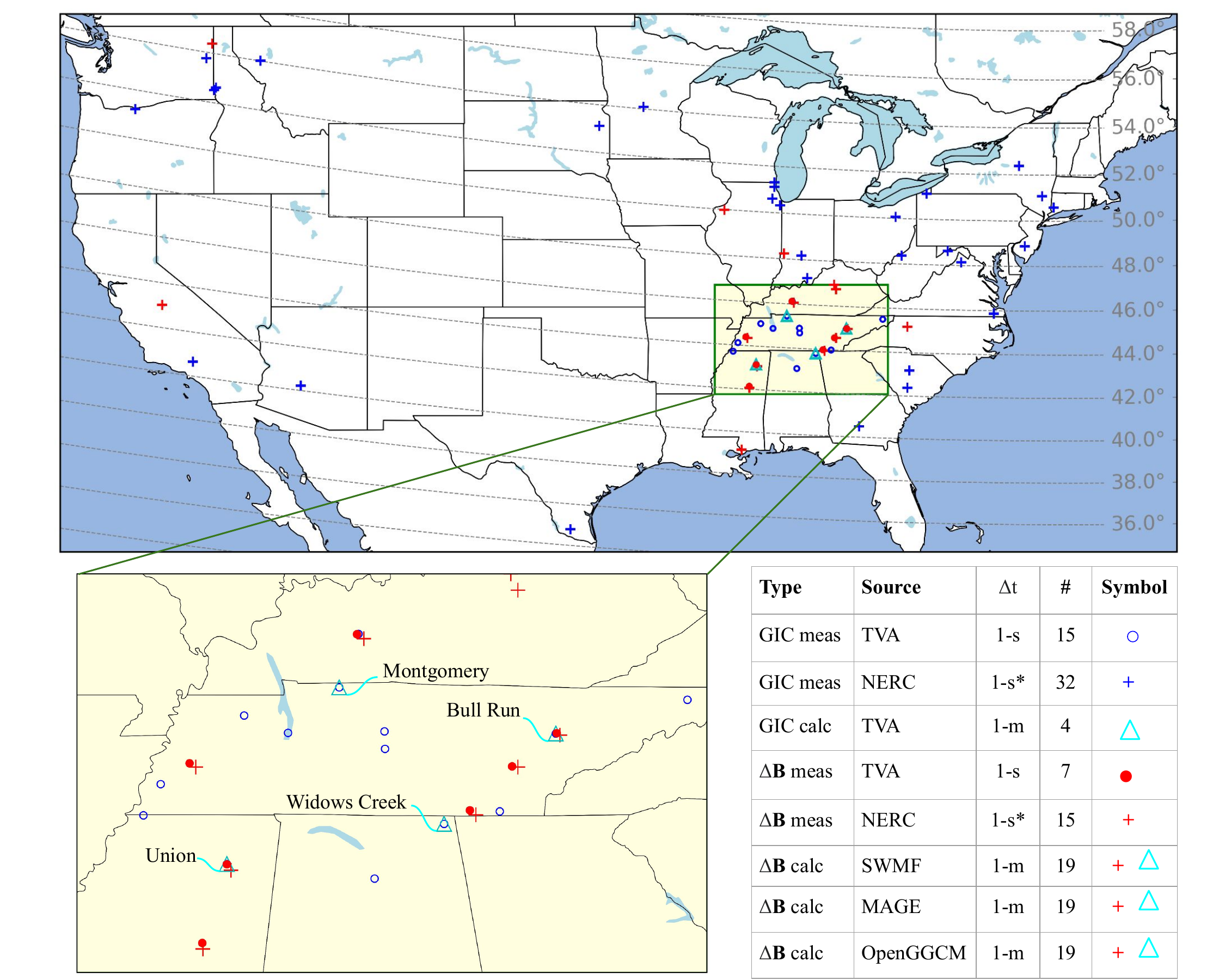}
    \label{fig:loc_map}
\end{figure}

\begin{figure}[h]
     \centering
     \begin{subfigure}[b]{0.49\textwidth}
         \centering
         \includegraphics[width=\textwidth]{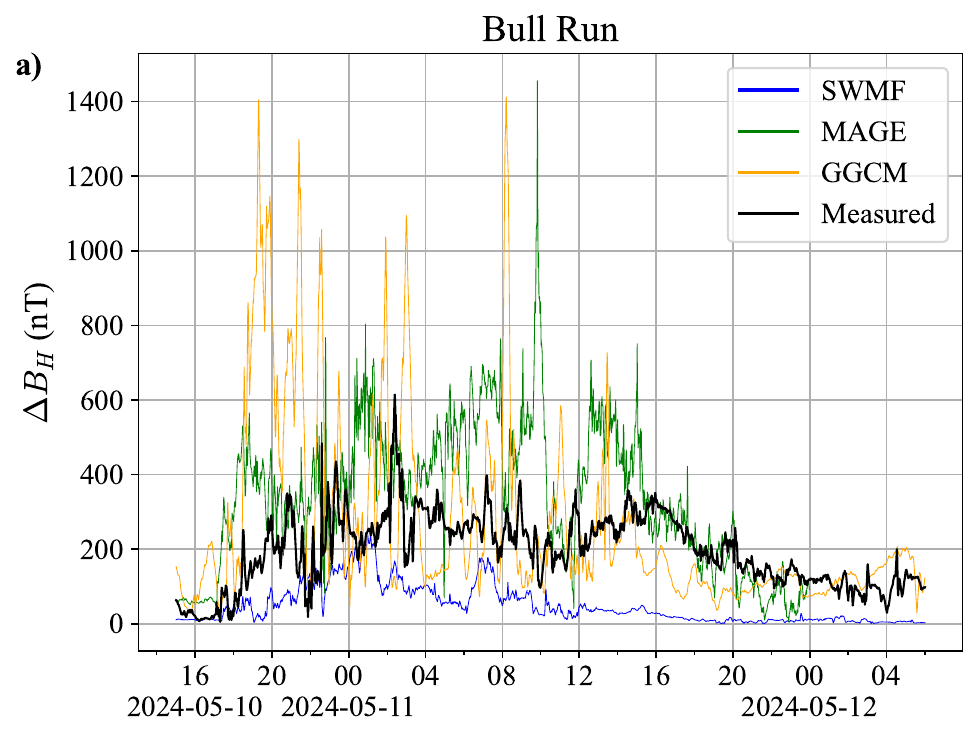}
         \label{br_b}
     \end{subfigure}
     \begin{subfigure}[b]{0.49\textwidth}
         \centering
         \includegraphics[width=\textwidth]{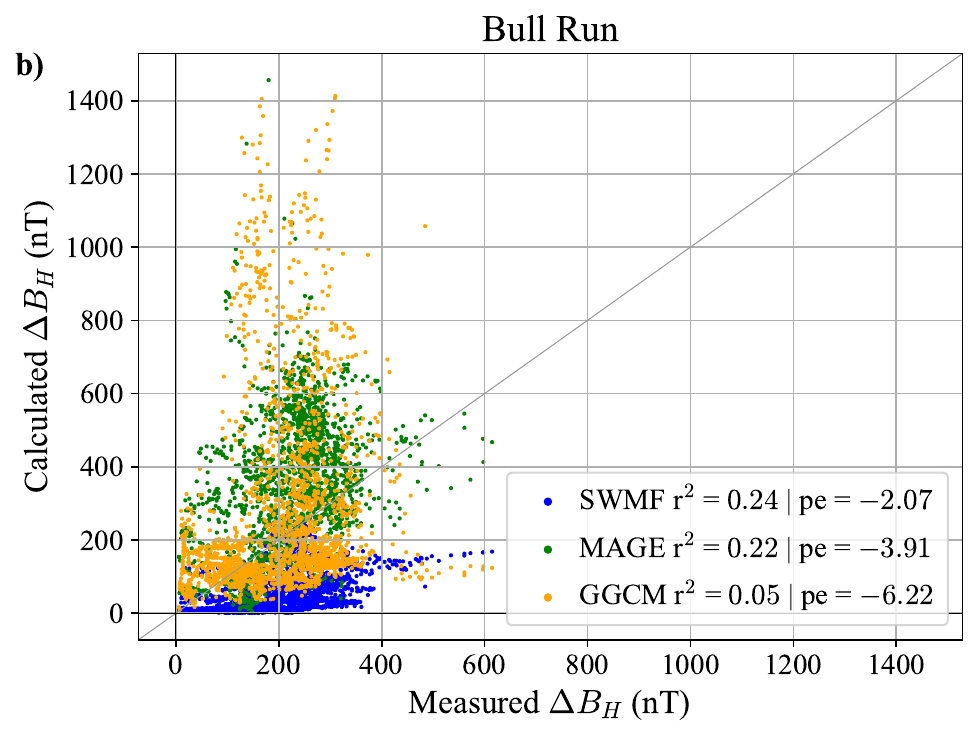}
         \label{br_corb}
     \end{subfigure}
     \begin{subfigure}[b]{0.49\textwidth}
         \centering
         \includegraphics[width=\textwidth]{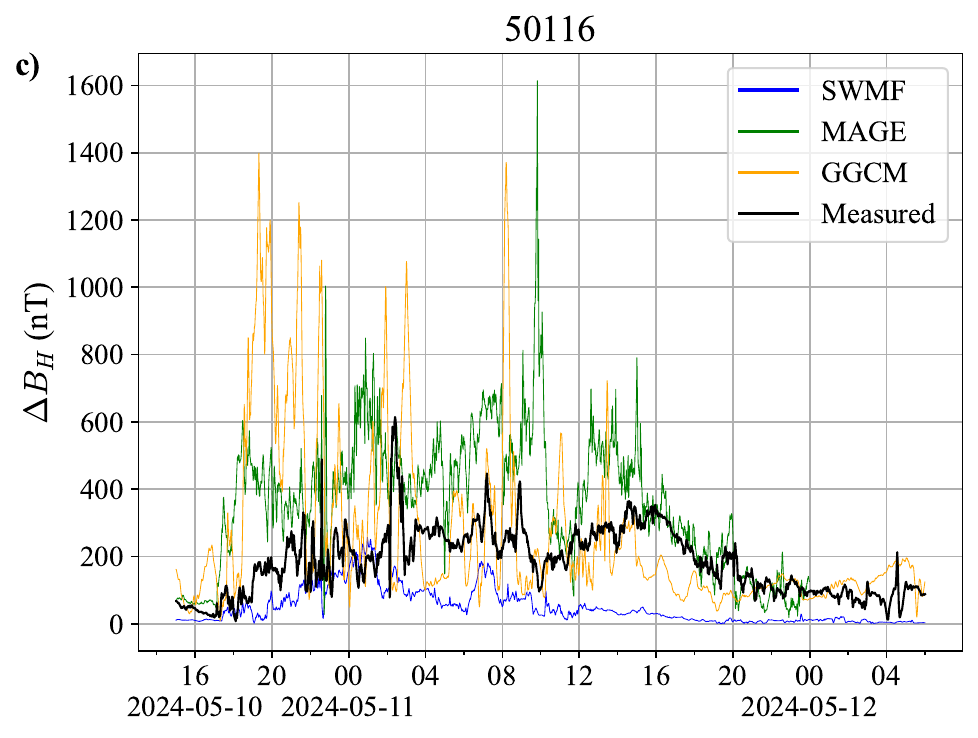}
         \label{nerc_b}
     \end{subfigure}
     \begin{subfigure}[b]{0.49\textwidth}
         \centering
         \includegraphics[width=\textwidth]{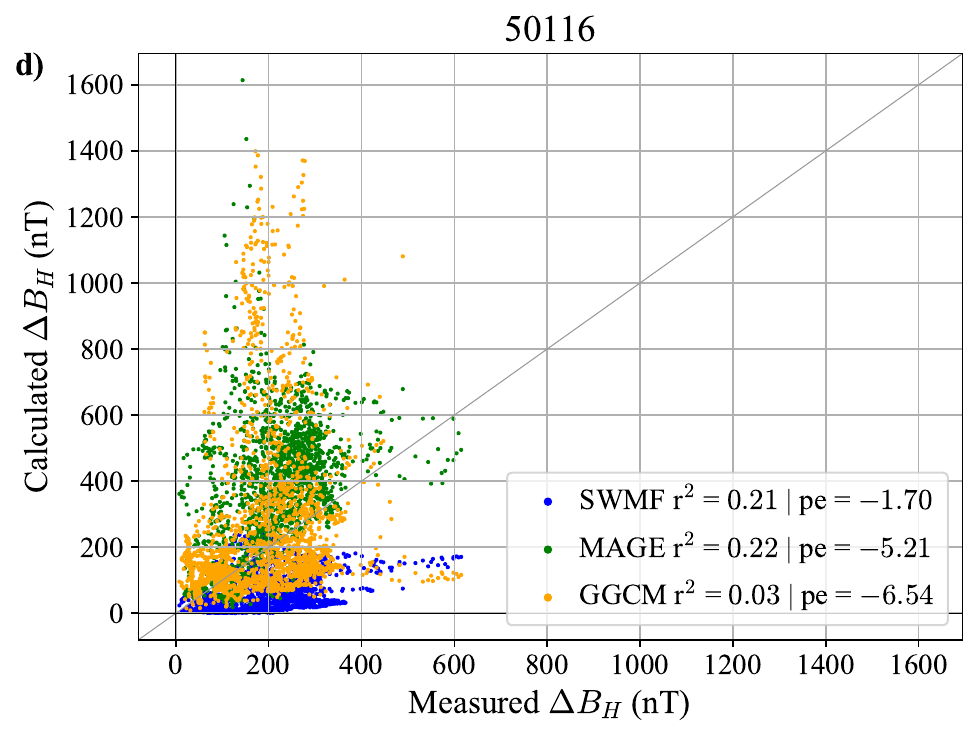}
         \label{nerc_corb}
     \end{subfigure}
        \caption{Calculated (SWMF, MAGE, and OpenGGCM) and measured $\Delta B_H$ timeseries comparison (left) and scatter correlation plots (right).}
        \label{fig:B}
    %\note[LW]{Updated all Fig 7 plots to have site names and timeseries plots (left) to have "Measured" on top.}
\end{figure}

The MAGE, SWMF, and OpenGGCM computed $\Delta B_H\equiv\sqrt{(\Delta B_x)^2+(\Delta B_y)^2}$ were compared to the measured $\Delta B_H$ for all 15 NERC magnetometer sites and to the 2 named TVA sites in Figure~\ref{fig:loc_map}, Bull Run and Union, that had co-located magnetometers. A representative result is shown in Figure~\ref{fig:B}. To compute $\mbox{r}$ and $\mbox{pe}$, the 1~s cadence difference in the surface magnetic field from a baseline ($\Delta \mathbf{B}$) measurements were averaged to 1~min, which was the cadence of the $\Delta \mathbf{B}$ model predictions.

Visual comparison of the model predictions to measured $\Delta B_H$ for 17 of the 22 magnetometer sites where simulation results were available from all three models in Figure~\ref{fig:B} shows a consistent result of MAGE over-predicting and SWMF under-predicting. The OpenGGCM $\Delta B_H$ predictions are generally between those of MAGE and SWMF. These observed data-model discrepancies are most likely a result of the many approximations required in MHD modeling \cite{Florczak2023}.
For all magnetometer sites, MAGE had a mean $\mbox{r}=0.44\pm 0.03$ and a mean $\mbox{pe} = -3.5\pm 0.7$. SWMF had a mean $\mbox{r}=0.58\pm 0.01$ and a mean $\mbox{pe} = -0.682\pm 0.07$. (The uncertainty is the standard error of the mean.) These results are consistent with the results in \citeA{Pulkkinen2011}, in which the average prediction efficiency of ground magnetic fields from global magnetosphere models computed at 12 northern hemisphere locations was, on average, less than zero for the four geomagnetic storm events considered. Additionally, \citeA{Welling2017} found that both SWMF and OpenGGCM have an overall tendency to under-predict.

The fact that all three models have non-zero data/model correlations indicates that they have predictive ability; the fact that the models consistently over- or under-predict suggests that a scaling factor could be applied to increase the prediction efficiencies if this behavior is found to be consistent across many storms. 
Note that the prediction efficiency is one of many possible prediction metrics. It is an overall prediction metric, and a model with a negative prediction efficiency can still have predictive ability in a restricted frequency range or in predicting threshold crossing events \cite{Pulkkinen2011}.

Prediction metrics for all sites where $\Delta \mathbf{B}$ measurements are available are shown in Table~\ref{table:all_sites_dB} of the Appendix.

These MHD simulation results come with \change[LW]{a caveat}{two caveats}. \change[LW]{In general}{First}, MHD simulations \add[LW]{in general }are complex and offer numerous options to choose from. These options include the inner magnetosphere model to use, whether to update the Earth's dipole tilt, etc. The SWMF and OpenGGCM runs, for example, were executed using the default options available on the CCMC website. Changing these options could change the simulation results such that the accuracy is improved. 
\add[LW]{Second, the MAGE and SWMF/OpenGGCM runs did not use identical solar wind conditions. As a result, we expect that some of the differences in the metrics shown in Figures~\protect\ref{fig:B}b and d may be attributed to differences in the solar wind driving conditions.
However, we find that our conclusions, particularly the general trend of negative prediction efficiencies and large differences between model results, are the same if we consider only the interval 2024-05-10 12:00 through 2024-05-11 12:00 when the solar wind driving conditions were more similar (see Table~\protect\ref{table:all_sites_dB_crop} in Appendix).}

\subsection{$\alpha$ and $\beta$ factors}
\label{sect:beta}

The NERC TPL--007 geomagnetic disturbance standard \cite{NERC2020a} requires that U.S. power system operators assess GIC vulnerability using a benchmark simulated electric field with components $x,y$ having amplitudes of

$E_x = E_{ox}\alpha(\lambda)\beta_x (\phi,\lambda)\quad\mbox{and}\quad E_y = E_{oy}\alpha(\lambda)\beta_y (\phi,\lambda)\thinspace,$

\noindent
where $\lambda$ and $\phi$ are geomagnetic latitude and longitude, respectively, $E_{0x}$ and $E_{0y}$ are 1-in-100-year peak geoelectric field amplitudes at a reference location, $\alpha=0.001e^{0.115\lambda}$ is a geomagnetic latitude scaling factor \protect\cite{NERC2020c}, and $\beta_x$ and $\beta_y$ are scaling factors that depend on the local ground conductivity as described in \citeA{Pulkkinen25}.

The components of $\boldsymbol{\beta}=[\beta_x, \beta_y]$ are computed by taking the ratio of the peak geoelectric field computed using a local ground conductivity model to the peak geoelectric field computed using a reference ground conductivity model. In NERC TPL--007, 1-D ground conductivity models were used to compute $\boldsymbol{\beta}$; \citeA{Pulkkinen25} used 3-D ground conductivity models derived from MT measurements at ${\sim}5,000$ sites on an approximately $0.5^\circ$ grid in the contiguous US to compute $\boldsymbol{\beta}$. In this work, we use these updated $\boldsymbol{\beta}$, described in Section~\ref{sect:intra}, because the ground conductivity models used to compute them are derived from measurements and have a much higher spatial resolution (there are only $19$ 1-D models for the contiguous US \cite{Fernberg2012}).

% This serves to define alpha and beta, but we don't use these electric fields in Dennis' or Chris' model. Also, we don't compare their fields with the benchmark field above.

\clearpage 

\clearpage

\clearpage

\begin{figure} [h]
    \caption{GIC times series from 15~TVA sites shown with 40~A baseline offsets, sorted by geomagnetic latitude, and labeled with the site name and its geomagnetic latitude and longitude in degrees. Data from nine sites that were also found in the NERC database are indicated with an asterisk.}
    \centering
    \includegraphics[width=1.0\textwidth]{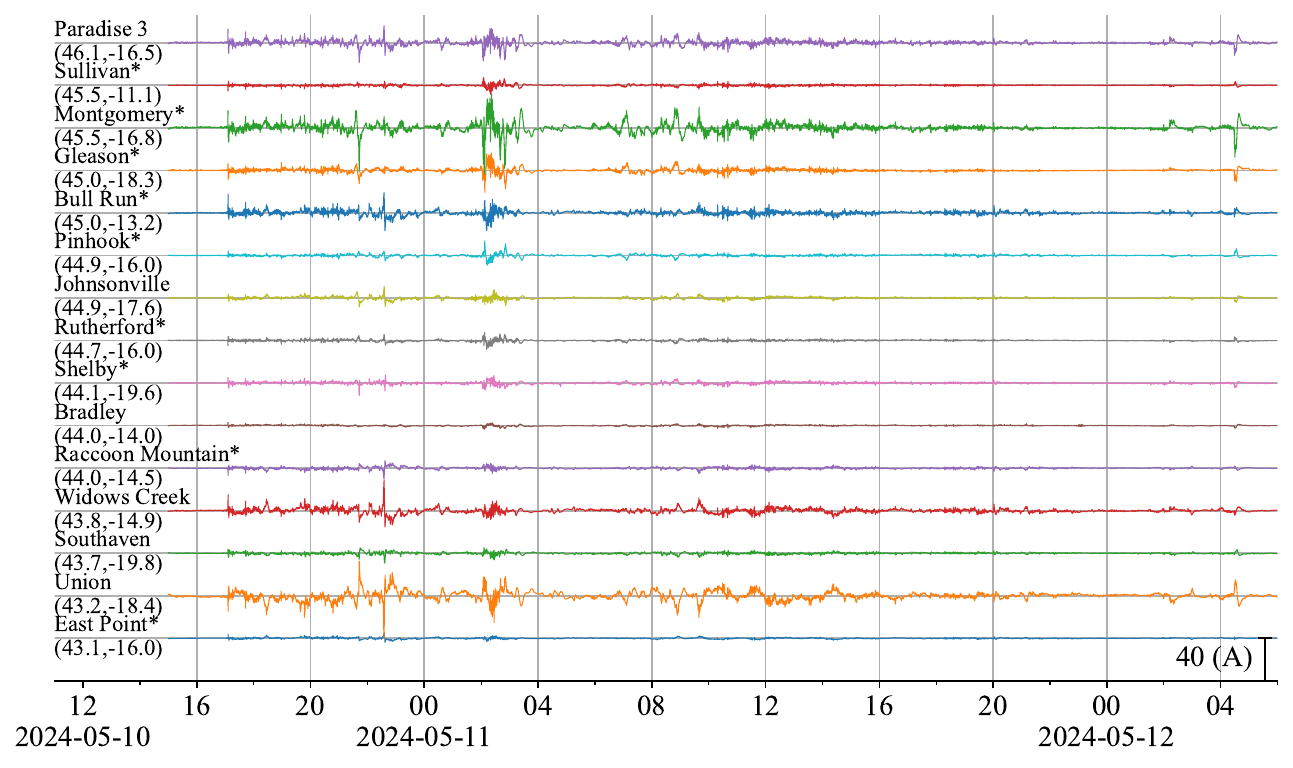}
    \label{fig:gic_tva}
    \note[LW]{Updated Fig 3 unit labels to be consistent.}
\end{figure}

\clearpage 

\begin{figure} [h]
    \caption{GIC data from 32~sites from NERC's ERO portal with 40~A baseline offsets sorted by geomagnetic latitude and labeled with the NERC site ID and its geomagnetic latitude and longitude in degrees. All NERC GIC data were either measured from a three-phase transformer or scaled to be equivalent.}
    \centering
    \includegraphics[width=1.0\textwidth]{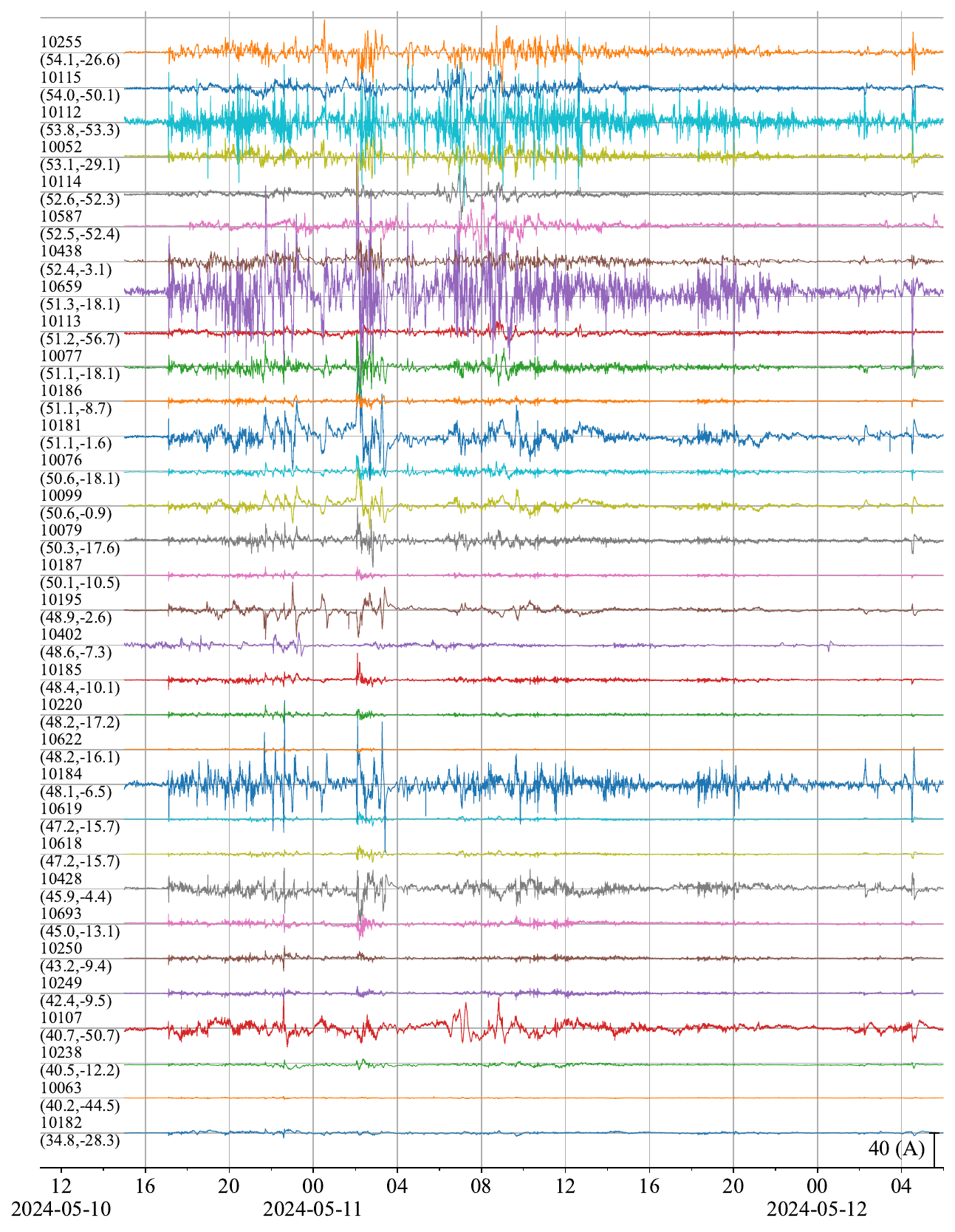}
    \label{fig:gic_nerc}
    \note[LW]{Updated Fig 4 unit labels to be consistent.}
\end{figure}

\clearpage

\begin{figure} [h]
    \caption{$\Delta B_H$ measurements from 15~sites from NERC's ERO portal and 7~sites from TVA with 400~nT baseline offsets sorted by geomagnetic latitude and labeled with the site ID and its geomagnetic latitude and longitude in degrees.}
    \centering
    \includegraphics[width=1.0\textwidth]{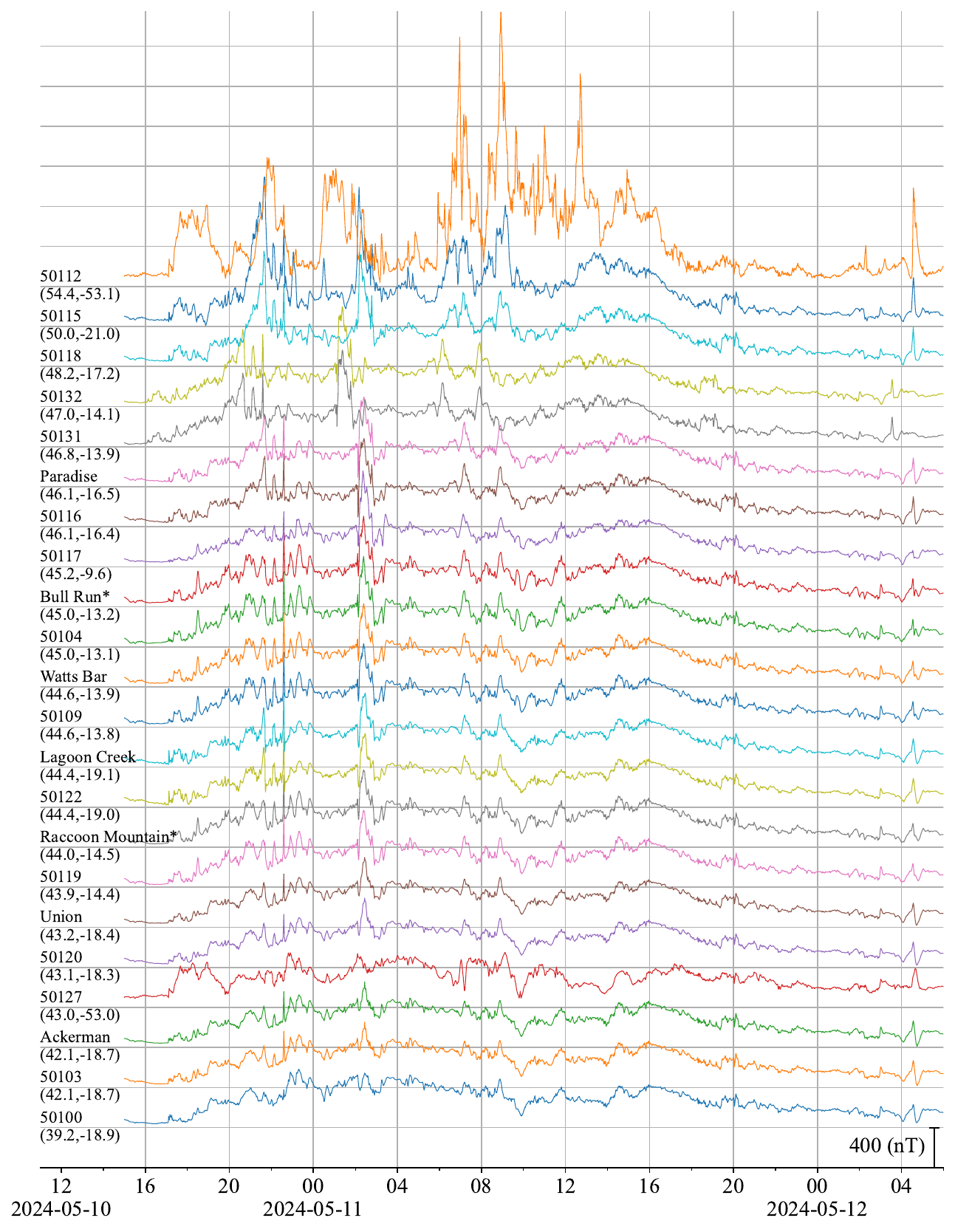}
    \label{fig:db_all}
    \note[LW]{Updated Fig 5 unit labels to be consistent.}
\end{figure}

\clearpage

\clearpage

\clearpage

\section{Empirical Relationships}

Ideally, GIC would be estimated by computing $\mathbf{B}(t)$ and a local ground conductivity model to determine $\mathbf{E}(t)$ and using this $\mathbf{E}(t)$ in a power system model with known parameters, as in the models described in Section~\ref{sect:gic-estimation}. However, real-time measured or high-quality forecasts of $\mathbf{B}(t)$ are not always available, and an accurate representation of the entire U.S. power system is not generally available, especially for historical events. As a result, empirical relationships that determine how summary measures of GIC at different locations are related to each other can be useful in their absence.

In Section~\ref{sect:intra}, we consider how GIC at a given site is correlated to GIC at other sites during the May 2024 geomagnetic storm. In Section~\ref{sect:max_std}, we consider how $|\mbox{GIC}|_{\text{max}}$ at a given site relates to geomagnetic latitude, $\lambda$, and the $\alpha$ and $\beta$ parameters described in Section~\ref{sect:beta}.

\subsection{Intra-site correlations}
\label{sect:intra}

In this section, we consider how measured GIC at the $47$ sites is related by comparing the GIC timeseries correlation between the $1,081$ unique site pairs. The pair correlations were related to the site pair separation distance, $\lambda$, $\beta$, the voltage of the power lines on which the GIC monitors were located, and the number of line connections.
The values of $\beta$ at each site were calculated by linearly interpolating $\beta$ values calculated at MT sites onto a $100\times100$ grid spanning all geographic longitudes and latitudes of the GIC and $\Delta \mathbf{B}$ sites (CONUS).

The $1,081$ site pairs vary in separation distance from $0$ to $3,957$~km. One of the pairs, sites 10619 and 10618, has zero separation distance due to the precision of latitude and longitude reported in the metadata ($0.1^\circ$); the next nearest pair has a separation of $5.19$~km. The highest $\beta$ values were along the eastern coast of the US and in the Great Lakes region, while the lowest were along the Gulf coast. 
% Power line voltage rating/category/level ; use highest operating voltage level line
The power line AC voltage on which the GIC monitors were mounted was estimated by finding the maximum line AC voltage, provided in the HIFLD dataset described in Section~\ref{sect:transmission-lines}, within a $500$~m radius of the projected latitude and longitude of the GIC monitor. Many transmission lines did not have an associated AC voltage in the HIFLD dataset. As a result, the magnitude of the voltage difference, $|\Delta V|$, was only estimated for $325$ of the $1,081$ site pairs. $43$\% of the pairs are associated with transmission lines of the same voltage (i.e., $|\Delta V| = 0$~kV). Statistics for separation distance, $|\Delta\lambda|$, $|\Delta\beta|$, and $|\Delta V|$ between site pairs are shown in Table \ref{table:compare_stats}.

\begin{figure}[h]
     \centering
     \begin{subfigure}[b]{0.49\textwidth}
         \centering
         \includegraphics[width=\textwidth]{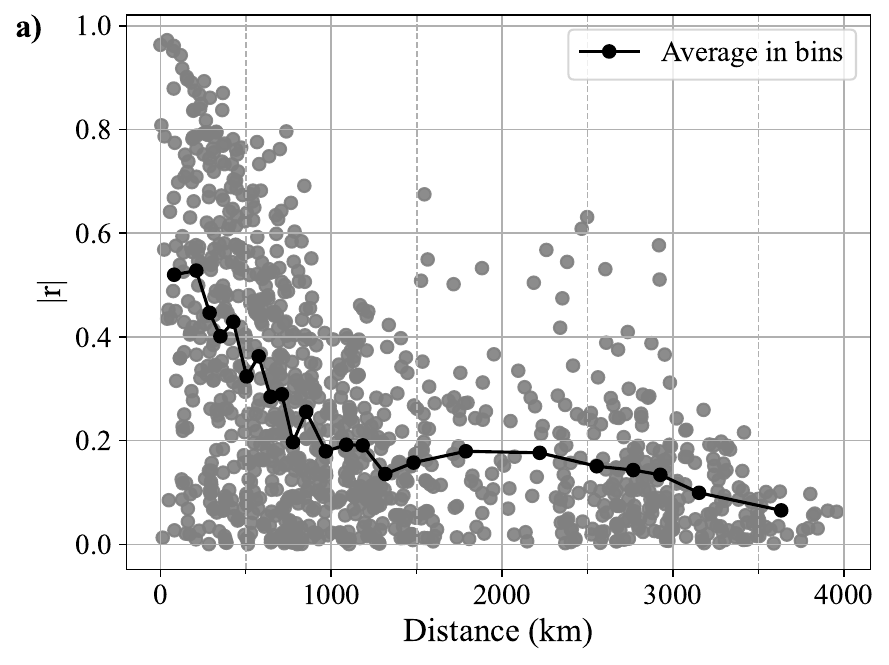}
         \label{dist}
     \end{subfigure}
     \begin{subfigure}[b]{0.49\textwidth}
         \centering
         \includegraphics[width=\textwidth]{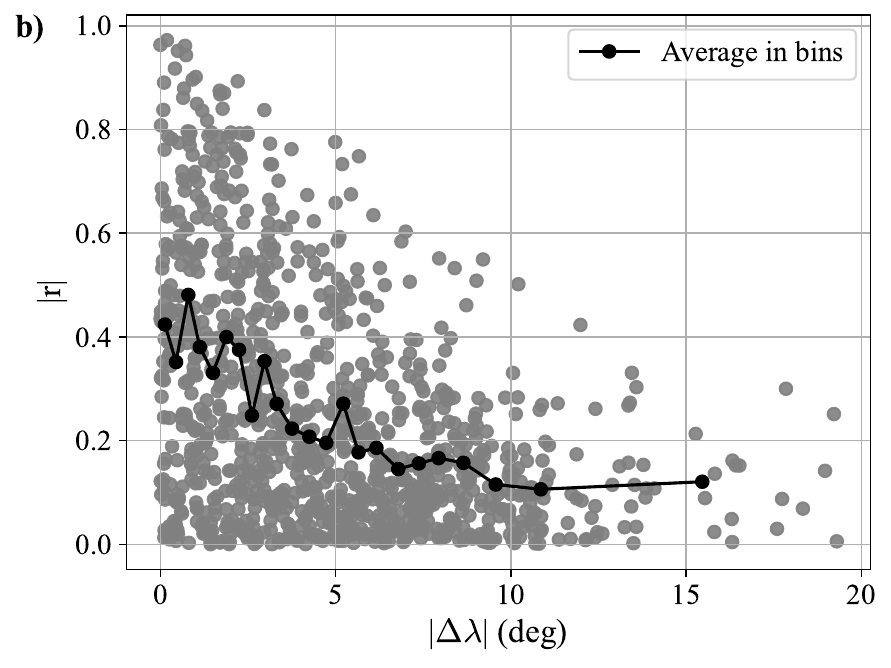}
         \label{beta}
     \end{subfigure}
     \begin{subfigure}[b]{0.49\textwidth}
         \centering
         \includegraphics[width=\textwidth]{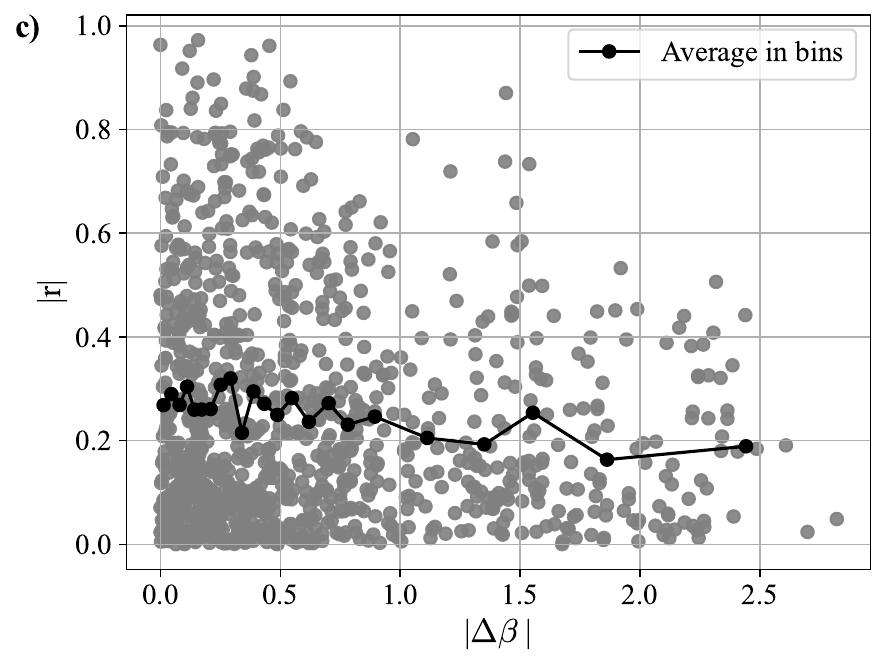}
         \label{volt}
     \end{subfigure}
     \begin{subfigure}[b]{0.49\textwidth}
         \centering
         \includegraphics[width=\textwidth]{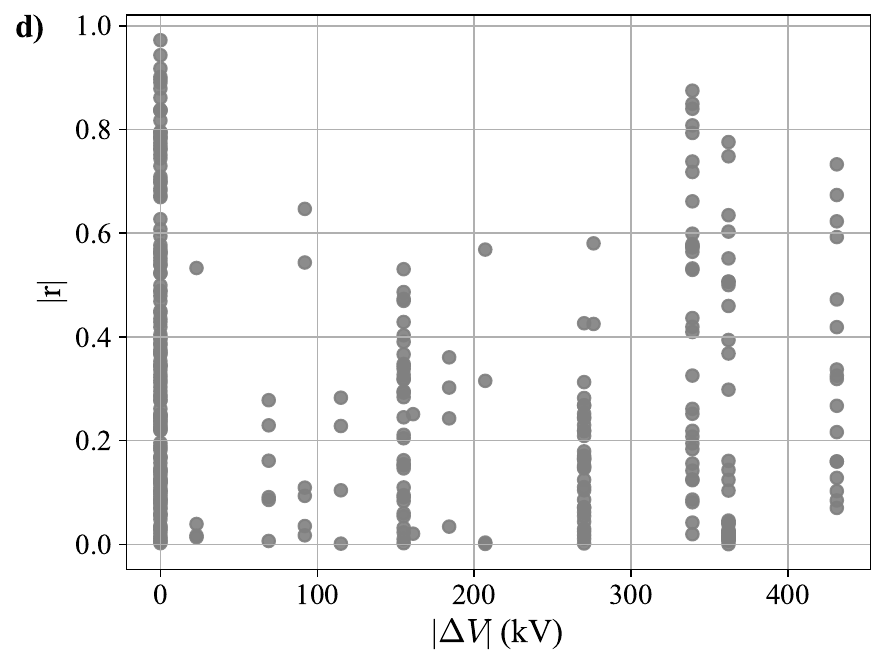}
         \label{lat}
     \end{subfigure}
        \caption{Correlation between GIC timeseries for $1,081$ site pairs (except for $\Delta V$ shown in subplot (d), which was only available for $325$ of the $1,081$ site pairs) vs. (a) differences in site pair separation distance, (b) geomagnetic latitude ($\lambda$), (c) conductivity scaling factor ($\beta$), and (d) power line voltage nearest to the GIC instrument. Mean $|$r$|$ values are shown as black dots centered on $23$ bins with $47$ points in each bin.}
        \label{fig:gic_scatter}
    %\note[LW]{Update Fig 8b x-axis label from $\Delta\lambda$ to $|\Delta\lambda|$.}
\end{figure}

\begin{figure}[h]
     \centering
     \begin{subfigure}[b]{0.49\textwidth}
         \centering
         \includegraphics[width=\textwidth]{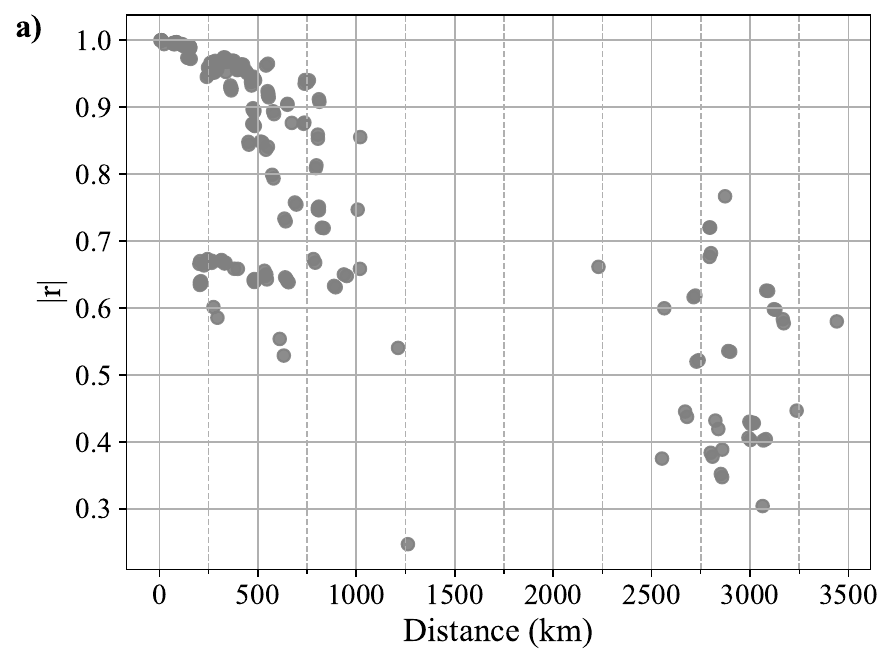}
         \label{dist}
     \end{subfigure}
     \begin{subfigure}[b]{0.49\textwidth}
         \centering
         \includegraphics[width=\textwidth]{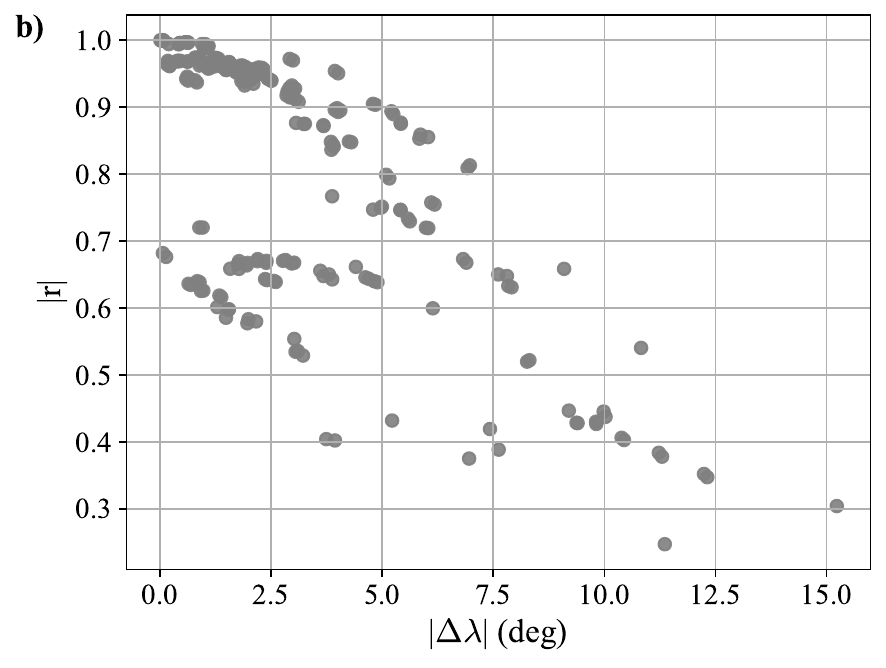}
         \label{beta}
     \end{subfigure}
     \begin{subfigure}[b]{0.49\textwidth}
         \centering
         \includegraphics[width=\textwidth]{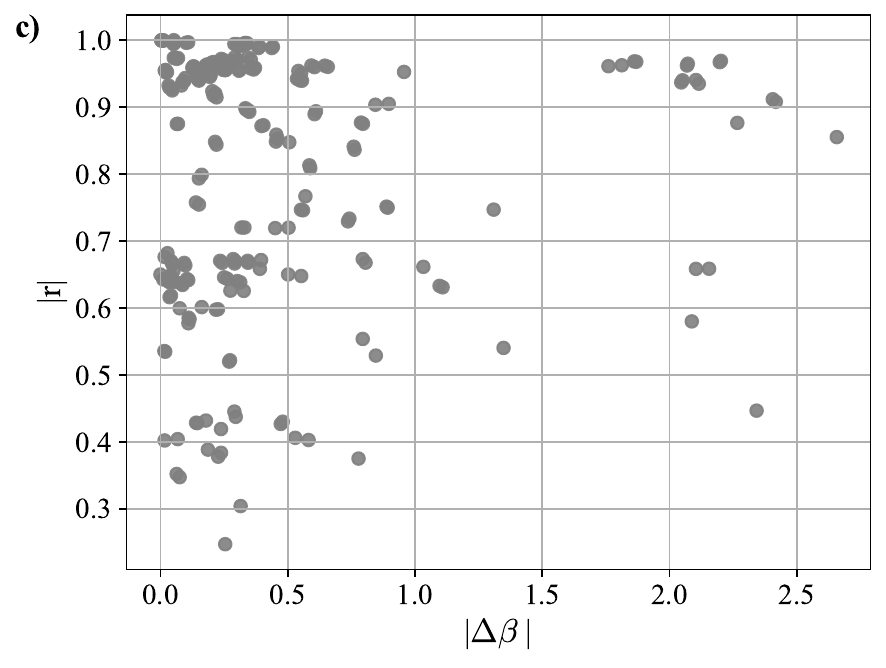}
         \label{volt}
     \end{subfigure}
     \begin{subfigure}[b]{0.49\textwidth}
         \centering
         \includegraphics[width=\textwidth]{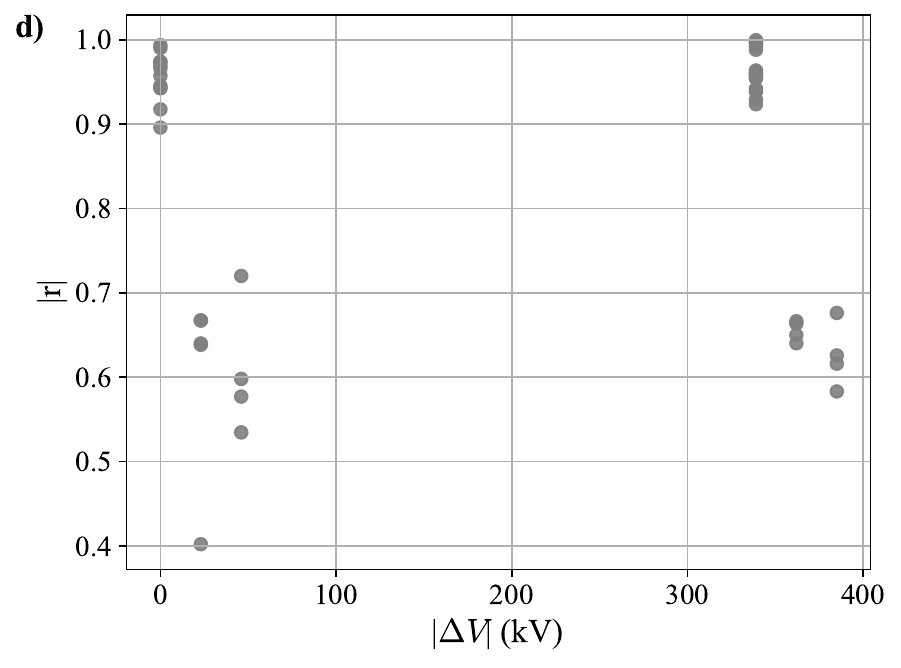}
         \label{lat}
     \end{subfigure}
        \vspace{-0.5in}
        \caption{Same as Figure~\ref{fig:gic_scatter} except for $\Delta B_H$ timeseries for all 231 magnetometer site pairs.}
        \label{fig:B_scatter}
    %\note[LW]{Update Fig 9b x-axis label from $\Delta\lambda$ to $|\Delta\lambda|$.}
\end{figure}

The results are shown in Figure~\ref{fig:gic_scatter}. The correlation between site pairs is weakly dependent on the site pair separation distance (Figure~\ref{fig:gic_scatter}a) and the difference in site geomagnetic latitude (Figure~\ref{fig:gic_scatter}b). Visually, only a weak relationship exists for differences in $\beta$ (Figure~\ref{fig:gic_scatter}c) and there is no relationship for the differences in nearest power line voltages (Figure~\ref{fig:gic_scatter}d).
The same analysis for $\Delta B_H$ timeseries measured at all 22 magnetometer sites (231 site pairs) is presented in Figure~\protect\ref{fig:B_scatter}.

These results, and the results of Section~\ref{sect:gic-estimation}, which showed GIC measured at four TVA sites is highly correlated with GIC computed using a power system network model, are consistent with the expectation that observed GIC is highly sensitive to network configuration and local ground conductivity. The correlation between waveforms in site pairs separated by less than $10$~km  (Figure~\ref{fig:gic_scatter}a) varies from near $0.0$ to near $1.0$. Similarly, site pairs with similar ground conductivity (based on only the proxy $\beta$) have $|$r$|$ from near $0.0$ to near $1.0$ (Figure~\ref{fig:gic_scatter}c). This conclusion can also be anticipated by comparing Figure~\ref{fig:db_all} with Figures~\ref{fig:gic_tva} and~\ref{fig:gic_nerc} --- the waveforms of $\Delta B_H$ are much more coherent than those of $\mbox{GIC}$. This can also be seen by comparing Figure~\ref{fig:B_scatter}a and b with Figure~\ref{fig:gic_scatter}a and b, which shows that the intra-site correlations in $\Delta B_H$ have a clearer dependency on distance and geomagnetic latitude than GIC.
However, the relationship between the inter-site correlations for $\Delta B_H$ with $|\Delta \beta|$ is weaker than that for GIC, which is consistent with the fact that GIC is more dependent on ground conductivity than $\Delta B_H$.
Furthermore, a similar intra-magnetometer comparison at high-latitude by \protect\citeA{Beggan2018} found that the confidence intervals of the mean-minute differences between the site pairs varied more along North-South separation than East-West. This result agrees with Figure~\ref{fig:B_scatter}b, which shows intra-magnetometer correlation decreasing with increased difference in geomagnetic latitude.

\begin{table}[h]
\centering
\begin{tabular}{lccc}
Variable & Min & Max & Mean \\
\hline
Distance (km) & 0.0 & 3,957 & 1,324 \\
$|\Delta\lambda|$ (deg) & 0.0 & 21.1 & 5.2 \\
$|\Delta\beta|$ & 0.0 & 1.57 & 0.29 \\
$|\Delta V|$ (kV) & 0.0 & 431 & 152 \\
\end{tabular}
\caption{Statistics of independent parameters used in Figure~\ref{fig:gic_scatter}.}
\label{table:compare_stats}
\end{table}

\clearpage

%\clearpage

\clearpage

\subsection{Maximum $|\mbox{GIC}|$}
\label{sect:max_std}

In the previous section, it was shown that the timeseries of GIC were only weakly correlated between sites, with an average correlation between all pairs of ${\sim}0.3$. 

In this section, we consider whether the maximum GIC magnitude, $|\mbox{GIC}|_{\text{max}}$, during the storm can be estimated using $\alpha$ and $\beta$. 
Predicting $|\mbox{GIC}|_{\text{max}}$ is of particular interest because power grids are most impacted by local peaks in the geoelectric field --- and therefore peaks in GIC --- and their durations \protect\cite{Abt2019}.
Additionally, using $\alpha$ and $\beta$ results in localized GIC estimates, which may be useful as power grid operators require more localized descriptions of storm severity \protect(\citeA{Abt2019}; \citeA{SWAG2024}).
To do this, we construct regression models of the form $|\mbox{GIC}|_{\text{max}} = a_\alpha\alpha + b_\alpha$, $|\mbox{GIC}|_{\text{max}} = a_\beta\beta + b_\beta$, and  $|\mbox{GIC}|_{\text{max}} = a_{\alpha\beta}\alpha\beta + b_{\alpha\beta}$ where $a$ and $b$ are regression parameters.The values of $\beta$ were interpolated at each site as described in Section~\protect\ref{sect:intra}, and the values of $\alpha$ were calculated directly using $\alpha=0.001e^{0.115\lambda}$, where $\lambda$ is the geomagnetic latitude of the site as described in Section \ref{sect:beta}.

\begin{figure}[h]
     \centering
     \begin{subfigure}[b]{0.7\textwidth}
         \centering
         \includegraphics[width=\textwidth]{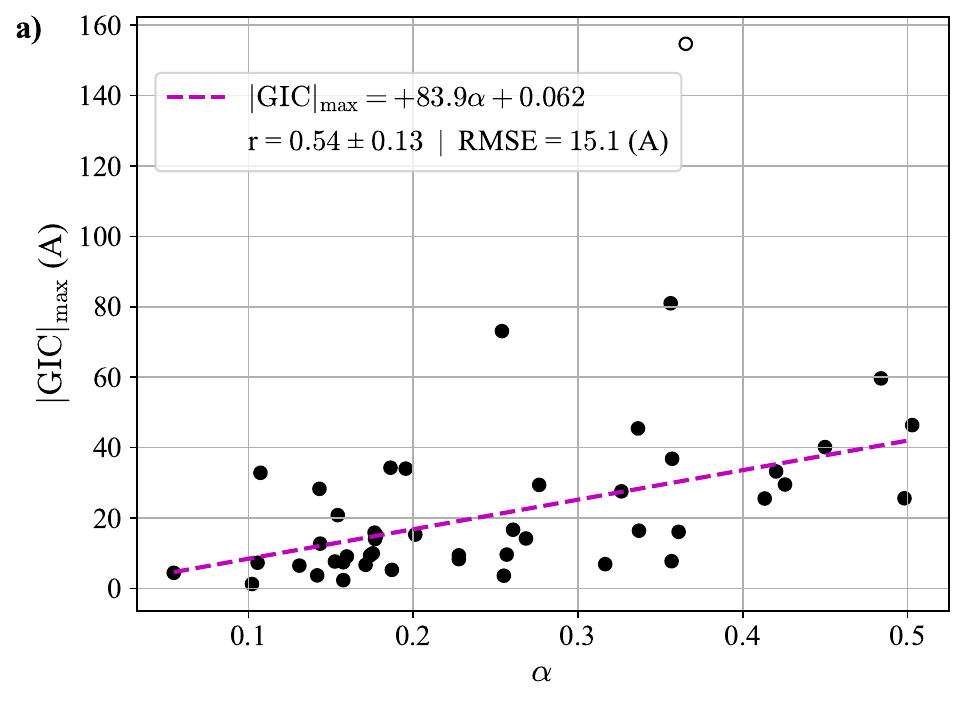}
         %\label{std}
     \end{subfigure}
     \begin{subfigure}[b]{0.7\textwidth}
         \centering
         \includegraphics[width=\textwidth]{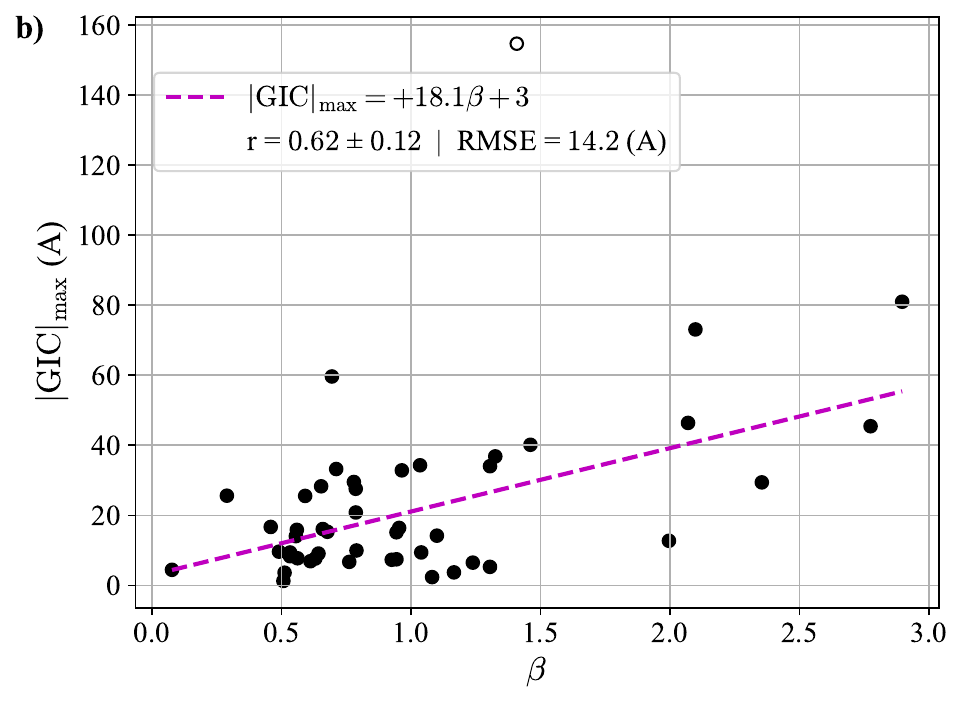}
         %\label{beta}
     \end{subfigure}
     \begin{subfigure}[b]{0.7\textwidth}
         \centering
         \includegraphics[width=\textwidth]{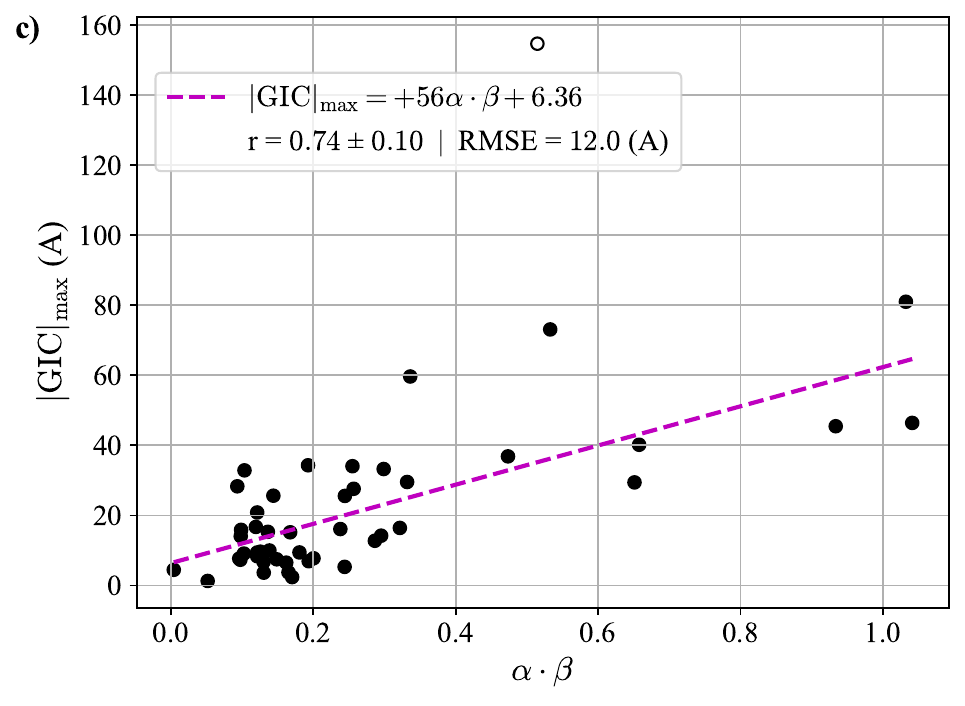}
         %\label{lat}
     \end{subfigure}
        \caption{GIC linear regression relationships. The white circle is an outlier that was excluded from the calculation of model parameters.}
        \label{fig:scatter2}
    \note[LW]{Updated Fig 10 unit labels to be consistent.}
\end{figure}

In Figure~\ref{fig:scatter2}a and b, we see that $|\mbox{GIC}|_{\text{max}}$ has a moderate linear relationship with $\alpha$ and $\beta$, respectively, with each model having $\mbox{r}\sim 0.6$ ($\mbox{r}^2\sim 0.36$). Visually, this relationship appears applicable for $44$ of the $47$ sites, as the sites with the three largest values of $|\mbox{GIC}|_{\text{max}}$ are significantly above the regression line. The root-mean-square error (RMSE) for each model is ${\sim}15$~A, which is a measure of uncertainty if the fit line was used to estimate $|\mbox{GIC}|_{\text{max}}$.

In Figure~\ref{fig:scatter2}c, the model that uses the product of $\alpha$ and $\beta$ has the highest r and lowest RMSE with respect to the linear models of Figure~\ref{fig:scatter2}a and b. 
However, according to the hierarchical principle, if $\alpha\beta$ is included as a parameter, all combinations of models using $\alpha$, $\beta$, and $\alpha\beta$ should be considered \cite{Bernhardt1979}. Furthermore, r and RMSE metrics do not take into account the number of parameters used in the regression model. Therefore, to compare models, we need a criterion that accounts for model complexity and the number of parameters.
The Akaike information criterion (AIC; \citeA{Burnham2004}) uses information loss as the criterion for identifying the best model. Using differences in AIC, one can compute the relative probability that one model minimizes information loss when it is used to represent the data compared to another model.

% See page 271 of https://sites.warnercnr.colostate.edu/wp-content/uploads/sites/73/2017/05/Burnham-and-Anderson-2004-SMR.pdf:

Table~\ref{table:regression_max} contains regression fit equations for all combinations of $\alpha$, $\beta$, and $\alpha\beta$, along with each fit's correlation coefficient $\pm$ twice its standard error (SE), $\mbox{r}^2$, RMSE, and AIC.
Using the rule described in \citeA{Burnham2004}, we can determine how much support there is for the possibility that a given fit minimizes the information loss relative to the fit with the minimum AIC value. 
This rule uses $\Delta_i=\mbox{AIC}-\mbox{AIC}_{\text{min}}$, the difference between the AIC value of model $i$ and the AIC value of the model with the minimum AIC. If $\Delta_i\le2$, then model $i$ is said to have ``substantial support" with respect to the model with the minimum AIC; if $4\le\Delta_i\le7$, then the selected model has ``considerably less support"; and, if $\Delta_i>10$, then the selected model has ``essentially no support".
Using this rule for the fits in Table~\ref{table:regression_max} with respect to Model~4 (which has the minimum AIC value), we find that Models~3--7 have $\Delta_i\le 2$, from which we conclude they have substantial support.
Models~1 and 2 have $\Delta_i>10$, corresponding to essentially no support.
This result is expected, as Models~1 and 2 only rely on $\alpha$ (assuming constant ground conductivity) or $\beta$ (assuming constant geomagnetic latitude), respectively, to predict $|\mbox{GIC}|_{\text{max}}$. Conversely, Models~3--7 use both $\alpha$ and $\beta$, accounting for the impacts of both geomagnetic latitude and ground conductivity on GIC.

Model~3, which contains only the product of $\alpha\beta$, is used in the NERC TPL--007 geomagnetic disturbance standard \cite{NERC2020a}, as described in Section~\ref{sect:beta}. The results in Table~\ref{table:regression_max} indicate that Models~3--7 are equivalent in the sense that they all provide similar metrics.

The coefficients of the regression models are expected to depend on storm intensity, so Models~3--7 are not generally expected to be useful for $|\mbox{GIC}|_{\text{max}}$ estimation unless the dependence of the coefficients on storm intensity is determined. (We have performed the same analysis as above for the 10 October 2024 storm and also found a linear relationship, but with a slope parameter for $\alpha\beta$ to be about one-half of that found here.) What we have shown in this work is that proportionality holds for a given storm, which is consistent with the assumption described in Section~\ref{sect:beta} that the maximum geoelectric field can be reasonably approximated for hazard assessment purposes using the product of $\alpha$ and $\beta$.

\clearpage

\begin{table}
\begin{tabular}{l l c c c c}
 & Fit Equation & r $\pm$ 2SE & r$^2$ & RMSE (A) & AIC \\
\hline
1 & $\vert{\text{GIC}\vert_\text{max}} = +88.9 \alpha  -1.67$ & $0.57 \pm 0.12$ & $0.33$ & $14.7$ & $381.9$ \\
2 & $\vert{\text{GIC}\vert_\text{max}} = +18.1 \beta  +2.48$ & $0.62 \pm 0.12$ & $0.38$ & $14.1$ & $378.2$ \\
3 & $\vert{\text{GIC}\vert_\text{max}} = +57.3 \alpha \cdot \beta  +5.54$ & $0.76 \pm 0.10$ & $0.58$ & $11.6$ & $360.2$ \\
4 & $\vert{\text{GIC}\vert_\text{max}} = +30.5 \alpha +48.7 \alpha \cdot \beta  +0.178$ & $0.78 \pm 0.09$ & $0.61$ & $11.3$ & $359.3$ \\
5 & $\vert{\text{GIC}\vert_\text{max}} = -3.97 \beta +66 \alpha \cdot \beta  +7.22$ & $0.77 \pm 0.10$ & $0.59$ & $11.6$ & $361.7$ \\
6 & $\vert{\text{GIC}\vert_\text{max}} = +74.1 \alpha +15.6 \beta  -13.5$ & $0.78 \pm 0.09$ & $0.60$ & $11.3$ & $359.7$ \\
7 & $\vert{\text{GIC}\vert_\text{max}} = +48.9 \alpha +7.2 \beta +27.7 \alpha \cdot \beta  -6.09$ & $0.78 \pm 0.09$ & $0.61$ & $11.2$ & $360.5$ \\
\end{tabular}

\caption{Regression equations and associated metrics considered for $|\mbox{GIC}|_{\text{max}}$. The model equations and data for fits 1, 2, and 3 --- which can be easily represented in a 2-D plot --- are shown in Figure~\ref{fig:scatter2}.}
\label{table:regression_max}
\end{table}

\clearpage

\section{Summary and Conclusions}

In this work, we have analyzed GIC-related measurements and model calculations and showed that network GIC models (from measured ground magnetic fields convolved with an impedance tensor to produce a local geoelectric field, then used to drive a power system simulation) produce predictions of measured GIC with correlations greater than 0.8 and prediction efficiencies in the range of $0.4$ to $0.7$. The Reference model, which uses estimates for unknown power system network parameters and the same geoelectric field, gives lower prediction efficiencies, as expected. The difference between the power system operator-computed results and the Reference model can be used to estimate uncertainty when only the Reference model results are available.

The results from global magnetosphere models of $\Delta B_H$ were compared to related terrestrial measurements. Although the measurement/model correlations of $\Delta B_H$ are ${\sim}0.5$, the prediction efficiencies are, on average, negative, indicating that the predicted fraction of the variance of the observed $\Delta B_H$ is small. This result indicates that, if model-predicted $\Delta B_H$ were used in place of measured $\Delta B_H$ to compute the geoelectric field, the quality of the $\mbox{GIC}$ predictions will be significantly lower than when measured $\Delta B_H$ is used.

The measured $\Delta B_H$ timeseries is more coherent across large distances than $\mbox{GIC}$. The correlation between measured timeseries at sites within $100$~km varies between $0.01$ and $0.97$ for $\mbox{GIC}$ and between $0.99$ and $1.0$ for $\Delta B_H$. 
This result confirms the expectation that power system network geometry and parameters have a significant influence on $\mbox{GIC}$, even if the local magnetic and electric field is similar between two sites.

We have found that the maximum magnitude of GIC during this storm can be estimated using a linear relationship with a slope of $\alpha\beta$, where $\alpha$ depends on geomagnetic latitude and $\beta$ is a scaling factor that depends on local ground conductivity. 
This linear relationship is consistent with the assumption of~\citeA{NERC2020a}, which recommends using a benchmark geoelectric field proportional to $\alpha\beta$ for power system vulnerability assessment.
Despite the limitations of this model, it can be used as a benchmark to model $|\mbox{GIC}|_{\mbox{max}}$ during the May 2024 Gannon storm without needing magnetometer data or network information.

\section*{Data Availability Statement}
All data used in this manuscript, including solar wind condition inputs for the SWMF, MAGE, and OpenGGCM models, are available at a GitHub data repository \cite{SWERVErepo2026} with the exception of NRCan and USGS magnetometer data, which can be downloaded from the USGS Geomagnetism Program (\url{https://geomag.usgs.gov}; \citeA{USGS1985}) and the NRCan Geomagnetism Program (\url{http://www.geomag.nrcan.gc.ca/}; \citeA{Newitt2007}) and from 1991 to present from INTERMAGNET (\url{https://www.intermagnet.org}; \citeA{Love2013}).

\acknowledgments
We acknowledge data provided directly by TVA and data from the NERC ERO Portal \cite{NERC24} and its data providers, which include contributions from the Electric Power Research Institute (EPRI) SUNBURST network. This work is supported by funding from NSF RAPID grant (\#2434136) and the NSF National Center for Atmospheric Research (\#1852977) via the Early-Career Faculty Innovator Program. Additionally, MW and the MAGE model development are supported by the NASA DRIVE Science Center for Geospace Storms (CGS) under award 80NSSC22M0163. We also acknowledge the USGS and NRCan for funding, operating, maintaining, and providing the operational magnetometer data that were used in this study. We thank the funding agencies and investigators who have openly provided their magnetotelluric transfer function products, specifically USArray (NSF), USMTArray (USGS and NASA).

The Community Coordinated Modeling Center (CCMC) executed the SWMF and OpenGGCM simulations at Goddard Space Flight Center (Run ID Dean\_Thomas\_070824\_3).

\section*{Conflict of Interest}
The authors declare that this research was conducted in the absence of any commercial or financial relationships that could be construed as a potential conflict of interest.

\clearpage

\section{Appendix}

\begin{figure}
    \centering
    \includegraphics[width=1.0\linewidth]{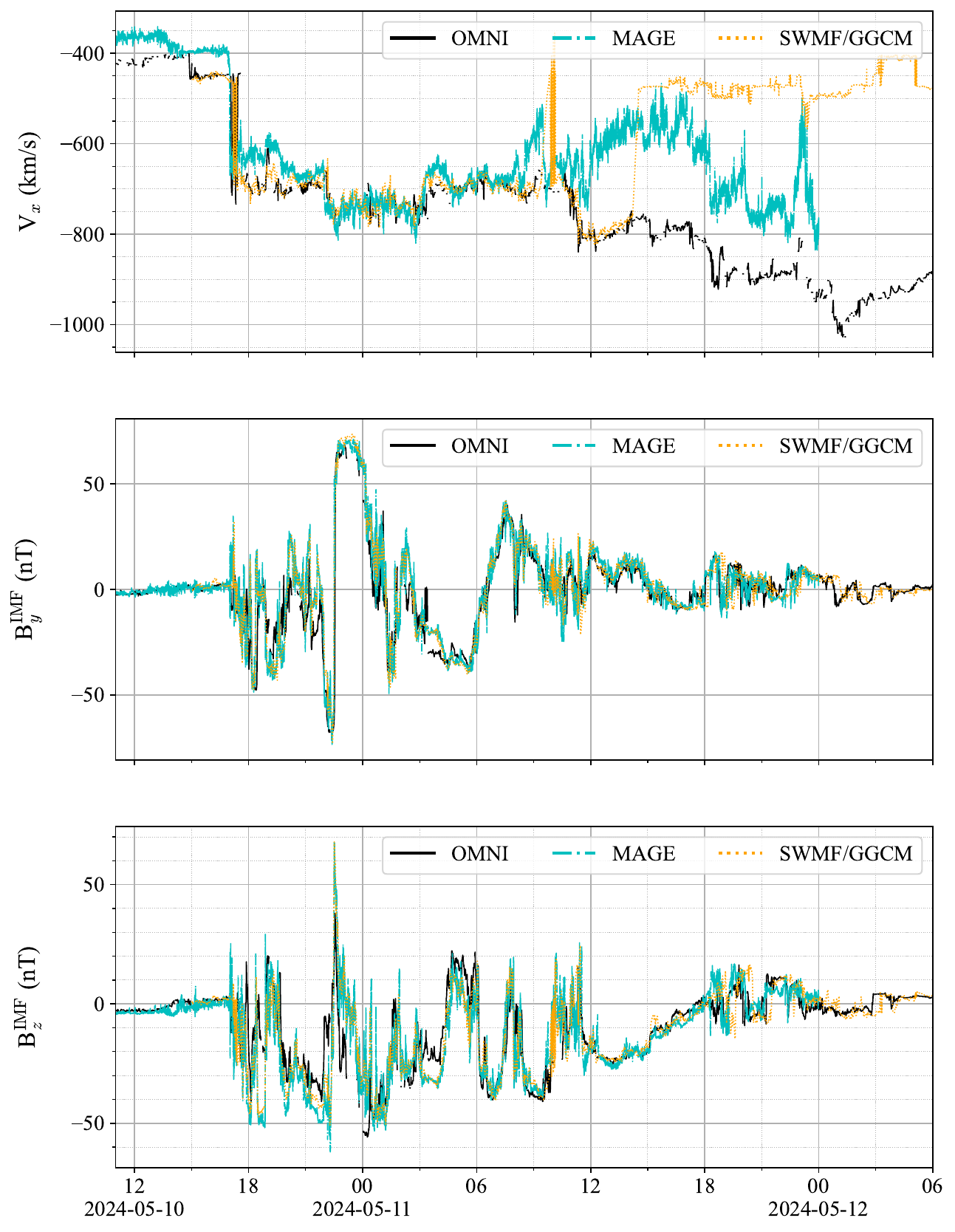}
    \caption{Comparison of V$_x$, B$_y^\text{IMF}$, and B$_z^\text{IMF}$ used as inputs for the different geospace environment models (MAGE, SWMF, OpenGGCM) as discussed in Section~\ref{sect:delta-b}. OMNI data is the same as that presented in Figure~\protect\ref{fig:solar_wind}.}
    \label{fig:compare_solar_wind}
    \note[LW]{Added figure for additional context on geospace environment model inputs.}
\end{figure}

\begingroup
\linespread{0.7}
\begin{table}[h]
\begin{tabular}{l r r r r r r r}
Site ID & $\sigma$ (A) & $\sigma_\text{TVA}$ & $\sigma_\text{Ref}$ & $\text{r}^2_\text{TVA}$ & $\text{r}^2_\text{Ref}$ & $\text{pe}_\text{TVA}$ & $\text{pe}_\text{Ref}$ \\
\hline
10052 & 4.2 &  & 4.3 &  & $0.47\phantom{*}$ &  & $0.25\phantom{*}$ \\
10076 & 1.8 &  & 0.5 &  & $0.14\phantom{*}$ &  & $-0.31*$ \\
10099 & 4.6 &  & 31.0 &  & $0.50\phantom{*}$ &  & $-36.77\phantom{*}$ \\
10238 & 1.0 &  & 2.9 &  & $0.01\phantom{*}$ &  & $-12.36\phantom{*}$ \\
10255 & 5.4 &  & 7.2 &  & $0.23\phantom{*}$ &  & $-0.49*$ \\
10587 & 3.6 &  & 5.1 &  & $0.00\phantom{*}$ &  & $-2.12*$ \\
10622 & 0.5 &  & 1.2 &  & $0.32\phantom{*}$ &  & $-4.98\phantom{*}$ \\
10063 & 0.2 &  & 0.3 &  & $0.08\phantom{*}$ &  & $-1.34*$ \\
10077 & 4.0 &  & 0.6 &  & $0.00\phantom{*}$ &  & $-0.15*$ \\
10079 & 2.9 &  & 0.5 &  & $0.28\phantom{*}$ &  & $0.17\phantom{*}$ \\
10113 & 2.1 &  & 1.4 &  & $0.16\phantom{*}$ &  & $-0.48*$ \\
10115 & 3.9 &  & 4.5 &  & $0.72\phantom{*}$ &  & $0.59*$ \\
10402 & 1.6 &  & 2.1 &  & $0.00\phantom{*}$ &  & $-1.88*$ \\
10428 & 4.6 &  & 3.2 &  & $0.43\phantom{*}$ &  & $0.43\phantom{*}$ \\
10659 & 18.0 &  & 0.6 &  & $0.15\phantom{*}$ &  & $-0.07\phantom{*}$ \\
10693 & 1.4 &  & 1.1 &  & $0.33\phantom{*}$ &  & $0.25\phantom{*}$ \\
10184 & 8.2 &  & 6.7 &  & $0.38\phantom{*}$ &  & $0.34\phantom{*}$ \\
10185 & 1.3 &  & 5.5 &  & $0.49\phantom{*}$ &  & $-12.49*$ \\
10186 & 1.1 &  & 0.2 &  & $0.31\phantom{*}$ &  & $0.15*$ \\
10187 & 0.7 &  & 0.6 &  & $0.25\phantom{*}$ &  & $0.10*$ \\
10195 & 3.9 &  & 5.4 &  & $0.47\phantom{*}$ &  & $-0.03*$ \\
10220 & 0.9 &  & 0.1 &  & $0.03\phantom{*}$ &  & $0.03*$ \\
10250 & 1.1 &  & 1.5 &  & $0.03\phantom{*}$ &  & $-1.63\phantom{*}$ \\
Bull Run & 1.5 & 2.2 & 1.1 & $0.80\phantom{*}$ & $0.48\phantom{*}$ & $0.50*$ & $0.48\phantom{*}$ \\
Gleason & 1.5 &  & 6.6 &  & $0.00\phantom{*}$ &  & $-18.64*$ \\
Johnsonville & 1.0 &  & 2.5 &  & $0.39\phantom{*}$ &  & $-3.56\phantom{*}$ \\
Montgomery & 3.7 & 1.0 & 1.3 & $0.86\phantom{*}$ & $0.15\phantom{*}$ & $0.43*$ & $0.15*$ \\
Paradise 3 & 2.0 &  & 5.3 &  & $0.21\phantom{*}$ &  & $-4.64\phantom{*}$ \\
Pinhook & 0.9 &  & 2.1 &  & $0.14\phantom{*}$ &  & $-3.32*$ \\
Raccoon Mountain & 0.8 &  & 0.6 &  & $0.51\phantom{*}$ &  & $0.50\phantom{*}$ \\
Rutherford & 0.7 &  & 1.0 &  & $0.07\phantom{*}$ &  & $-1.73\phantom{*}$ \\
Shelby & 0.8 &  & 1.1 &  & $0.01\phantom{*}$ &  & $-1.31\phantom{*}$ \\
Southaven & 0.8 &  & 0.3 &  & $0.19\phantom{*}$ &  & $0.19\phantom{*}$ \\
Sullivan & 0.6 &  & 0.9 &  & $0.21\phantom{*}$ &  & $-0.87\phantom{*}$ \\
Union & 3.8 & 3.5 & 1.8 & $0.67\phantom{*}$ & $0.57\phantom{*}$ & $0.65*$ & $0.50*$ \\
Widows Creek & 1.9 & 1.2 & 0.6 & $0.67\phantom{*}$ & $0.28\phantom{*}$ & $0.63*$ & $0.23*$ \\
Weakley & 0.9 &  & 3.5 &  & $0.00\phantom{*}$ &  & $-14.97\phantom{*}$ \\
10203 & 0.8 &  & 0.0 &  & $0.05\phantom{*}$ &  & $0.01\phantom{*}$ \\
10204 & 0.8 &  & 1.1 &  & $0.01\phantom{*}$ &  & $-1.31\phantom{*}$ \\
10660 & 1.5 &  & 6.6 &  & $0.00\phantom{*}$ &  & $-18.63*$ \\
10197 & 0.6 &  & 0.9 &  & $0.21\phantom{*}$ &  & $-0.87\phantom{*}$ \\
10208 & 0.7 &  & 1.0 &  & $0.07\phantom{*}$ &  & $-1.73\phantom{*}$ \\
10212 & 0.9 &  & 2.1 &  & $0.14\phantom{*}$ &  & $-3.31*$ \\
\hline
Mean & 2.4 & 2.0 & 3.0 & 0.75 & 0.22 & 0.55 & -3.39 \\
\end{tabular}

\vspace{-0.1in}
\caption{Prediction metrics for TVA and Reference model. When r is negative, the values of pe are computed by multiplying the measured GIC by $-1$, which assumes the GIC sensors are installed in an orientation opposite to that assumed by the models \protect\cite{Balch24}. These cases are indicated
with a *.}
%\note[LW]{Table has been updated with metrics from most recent Reference model run.}
\label{table:all_sites_GIC}
\end{table}
\endgroup

\begingroup
\linespread{0.8}
\begin{table}[h]
\begin{tabular}{l p{1cm} p{1cm} p{1cm} p{1cm} p{1cm} p{1cm} p{1cm} p{1cm} p{1cm} p{1cm}}
Site ID & $\sigma$ (nT) & $\sigma_\text{SWMF}$ & $\sigma_\text{MAGE}$ & $\sigma_\text{GGCM}$ & $\text{r}^2_\text{SWMF}$ & $\text{r}^2_\text{MAGE}$ & $\text{r}^2_\text{GGCM}$ & $\text{pe}_\text{SWMF}$ & $\text{pe}_\text{MAGE}$ & $\text{pe}_\text{GGCM}$ \\
\hline
50100 & 89.5 & 41.8 & 135.6 & 260.5 & $0.39\phantom{*}$ & $0.13\phantom{*}$ & $0.09\phantom{*}$ & $-2.42\phantom{*}$ & $-1.87\phantom{*}$ & $-7.63\phantom{*}$ \\
50127 & 71.4 & 42.3 & 125.5 & 244.4 & $0.13\phantom{*}$ & $0.14\phantom{*}$ & $0.05\phantom{*}$ & $-4.84\phantom{*}$ & $-2.35\phantom{*}$ & $-10.79\phantom{*}$ \\
50112 & 287.8 & 73.8 & 119.3 & 308.8 & $0.20\phantom{*}$ & $0.21\phantom{*}$ & $0.03\phantom{*}$ & $-0.53\phantom{*}$ & $0.12\phantom{*}$ & $-0.78\phantom{*}$ \\
50131 & 100.6 & 54.5 & 192.3 & 249.9 & $0.22\phantom{*}$ & $0.33\phantom{*}$ & $0.10\phantom{*}$ & $-1.38\phantom{*}$ & $-3.90\phantom{*}$ & $-5.15\phantom{*}$ \\
50132 & 104.0 & 54.9 & 192.7 & 248.4 & $0.24\phantom{*}$ & $0.35\phantom{*}$ & $0.10\phantom{*}$ & $-1.31\phantom{*}$ & $-3.32\phantom{*}$ & $-4.72\phantom{*}$ \\
50103 & 86.3 & 45.7 & 133.7 & 253.9 & $0.31\phantom{*}$ & $0.19\phantom{*}$ & $0.06\phantom{*}$ & $-2.20\phantom{*}$ & $-1.99\phantom{*}$ & $-8.26\phantom{*}$ \\
50104 & 95.8 & 51.3 & 189.8 & 252.4 & $0.25\phantom{*}$ & $0.23\phantom{*}$ & $0.05\phantom{*}$ & $-2.00\phantom{*}$ & $-4.00\phantom{*}$ & $-6.23\phantom{*}$ \\
50109 & 94.2 & 50.2 & 191.5 & 250.6 & $0.24\phantom{*}$ & $0.24\phantom{*}$ & $0.05\phantom{*}$ & $-1.92\phantom{*}$ & $-4.60\phantom{*}$ & $-6.48\phantom{*}$ \\
50115 & 167.7 & 61.1 & 128.3 & 264.6 & $0.16\phantom{*}$ & $0.30\phantom{*}$ & $0.04\phantom{*}$ & $-0.86\phantom{*}$ & $0.24\phantom{*}$ & $-1.88\phantom{*}$ \\
50116 & 97.2 & 52.8 & 197.1 & 253.8 & $0.21\phantom{*}$ & $0.22\phantom{*}$ & $0.03\phantom{*}$ & $-1.70\phantom{*}$ & $-5.21\phantom{*}$ & $-6.54\phantom{*}$ \\
50117 & 94.7 & 52.1 & 192.8 & 253.5 & $0.20\phantom{*}$ & $0.25\phantom{*}$ & $0.02\phantom{*}$ & $-1.61\phantom{*}$ & $-3.91\phantom{*}$ & $-7.26\phantom{*}$ \\
50118 & 121.0 & 57.2 & 197.2 & 253.8 & $0.16\phantom{*}$ & $0.21\phantom{*}$ & $0.03\phantom{*}$ & $-1.25\phantom{*}$ & $-2.69\phantom{*}$ & $-3.88\phantom{*}$ \\
50119 & 92.3 & 49.2 & 192.9 & 249.4 & $0.26\phantom{*}$ & $0.23\phantom{*}$ & $0.06\phantom{*}$ & $-1.88\phantom{*}$ & $-5.33\phantom{*}$ & $-6.73\phantom{*}$ \\
50120 & 86.6 & 47.3 & 198.7 & 250.3 & $0.26\phantom{*}$ & $0.19\phantom{*}$ & $0.04\phantom{*}$ & $-2.17\phantom{*}$ & $-8.04\phantom{*}$ & $-8.06\phantom{*}$ \\
50122 & 87.3 & 49.2 & 132.2 & 248.9 & $0.24\phantom{*}$ & $0.24\phantom{*}$ & $0.04\phantom{*}$ & $-2.03\phantom{*}$ & $-1.77\phantom{*}$ & $-7.90\phantom{*}$ \\
Bull Run & 96.1 & 51.3 & 189.9 & 252.5 & $0.24\phantom{*}$ & $0.22\phantom{*}$ & $0.05\phantom{*}$ & $-2.07\phantom{*}$ & $-3.91\phantom{*}$ & $-6.22\phantom{*}$ \\
Union & 87.7 & 47.4 & 198.9 & 250.0 & $0.27\phantom{*}$ & $0.20\phantom{*}$ & $0.05\phantom{*}$ & $-2.16\phantom{*}$ & $-7.75\phantom{*}$ & $-7.72\phantom{*}$ \\
\hline
Mean & 109.4 & 51.9 & 171.1 & 255.6 & 0.23 & 0.23 & 0.05 & -1.90 & -3.55 & -6.25 \\
\end{tabular}

\caption{Prediction metrics for SWMF, MAGE, and OpenGGCM for all sites where $\Delta\mathbf{B}$ measurements are available.}
\label{table:all_sites_dB}
\end{table}
\endgroup

\begingroup
\linespread{0.8}
\begin{table}[h]
\begin{tabular}{l p{1cm} p{1cm} p{1cm} p{1cm} p{1cm} p{1cm} p{1cm} p{1cm} p{1cm} p{1cm}}
Site ID & $\sigma$ (nT) & $\sigma_\text{SWMF}$ & $\sigma_\text{MAGE}$ & $\sigma_\text{GGCM}$ & $\text{r}^2_\text{SWMF}$ & $\text{r}^2_\text{MAGE}$ & $\text{r}^2_\text{GGCM}$ & $\text{pe}_\text{SWMF}$ & $\text{pe}_\text{MAGE}$ & $\text{pe}_\text{GGCM}$ \\
\hline
50100 & 113.1 & 45.5 & 170.4 & 301.8 & $0.47\phantom{*}$ & $0.31\phantom{*}$ & $0.09\phantom{*}$ & $-1.03\phantom{*}$ & $-0.66\phantom{*}$ & $-7.29\phantom{*}$ \\
50127 & 94.7 & 46.4 & 160.7 & 285.8 & $0.18\phantom{*}$ & $0.41\phantom{*}$ & $0.07\phantom{*}$ & $-2.13\phantom{*}$ & $-0.75\phantom{*}$ & $-8.94\phantom{*}$ \\
50112 & 314.3 & 81.7 & 155.4 & 355.4 & $0.16\phantom{*}$ & $0.35\phantom{*}$ & $0.01\phantom{*}$ & $-0.47\phantom{*}$ & $0.19\phantom{*}$ & $-1.05\phantom{*}$ \\
50131 & 109.3 & 59.5 & 213.1 & 287.5 & $0.27\phantom{*}$ & $0.38\phantom{*}$ & $0.09\phantom{*}$ & $-0.72\phantom{*}$ & $-3.67\phantom{*}$ & $-7.19\phantom{*}$ \\
50132 & 114.4 & 60.1 & 213.9 & 286.2 & $0.27\phantom{*}$ & $0.38\phantom{*}$ & $0.07\phantom{*}$ & $-0.71\phantom{*}$ & $-3.17\phantom{*}$ & $-6.45\phantom{*}$ \\
50103 & 104.2 & 50.0 & 168.7 & 292.2 & $0.43\phantom{*}$ & $0.35\phantom{*}$ & $0.07\phantom{*}$ & $-0.83\phantom{*}$ & $-0.88\phantom{*}$ & $-8.84\phantom{*}$ \\
50104 & 115.4 & 55.8 & 208.3 & 289.3 & $0.38\phantom{*}$ & $0.37\phantom{*}$ & $0.06\phantom{*}$ & $-0.67\phantom{*}$ & $-2.86\phantom{*}$ & $-6.81\phantom{*}$ \\
50109 & 110.2 & 54.7 & 212.9 & 287.5 & $0.37\phantom{*}$ & $0.37\phantom{*}$ & $0.06\phantom{*}$ & $-0.71\phantom{*}$ & $-3.60\phantom{*}$ & $-7.54\phantom{*}$ \\
50115 & 187.5 & 67.4 & 163.9 & 308.0 & $0.16\phantom{*}$ & $0.42\phantom{*}$ & $0.03\phantom{*}$ & $-0.51\phantom{*}$ & $0.37\phantom{*}$ & $-2.42\phantom{*}$ \\
50116 & 108.2 & 57.8 & 219.8 & 290.9 & $0.33\phantom{*}$ & $0.33\phantom{*}$ & $0.03\phantom{*}$ & $-0.66\phantom{*}$ & $-4.90\phantom{*}$ & $-8.52\phantom{*}$ \\
50117 & 106.6 & 56.4 & 197.0 & 290.5 & $0.34\phantom{*}$ & $0.32\phantom{*}$ & $0.02\phantom{*}$ & $-0.53\phantom{*}$ & $-3.47\phantom{*}$ & $-9.26\phantom{*}$ \\
50118 & 133.6 & 62.9 & 219.9 & 293.5 & $0.22\phantom{*}$ & $0.26\phantom{*}$ & $0.03\phantom{*}$ & $-0.57\phantom{*}$ & $-2.83\phantom{*}$ & $-5.18\phantom{*}$ \\
50119 & 107.8 & 53.6 & 215.6 & 286.7 & $0.39\phantom{*}$ & $0.37\phantom{*}$ & $0.06\phantom{*}$ & $-0.70\phantom{*}$ & $-4.13\phantom{*}$ & $-7.83\phantom{*}$ \\
50120 & 102.5 & 51.7 & 220.9 & 288.1 & $0.42\phantom{*}$ & $0.36\phantom{*}$ & $0.06\phantom{*}$ & $-0.74\phantom{*}$ & $-6.47\phantom{*}$ & $-9.01\phantom{*}$ \\
50122 & 100.6 & 54.0 & 167.4 & 286.1 & $0.39\phantom{*}$ & $0.39\phantom{*}$ & $0.05\phantom{*}$ & $-0.72\phantom{*}$ & $-0.95\phantom{*}$ & $-9.42\phantom{*}$ \\
Bull Run & 115.3 & 55.8 & 208.7 & 289.3 & $0.37\phantom{*}$ & $0.36\phantom{*}$ & $0.05\phantom{*}$ & $-0.72\phantom{*}$ & $-2.84\phantom{*}$ & $-6.82\phantom{*}$ \\
Union & 105.2 & 51.8 & 221.1 & 287.7 & $0.43\phantom{*}$ & $0.38\phantom{*}$ & $0.07\phantom{*}$ & $-0.69\phantom{*}$ & $-6.02\phantom{*}$ & $-8.39\phantom{*}$ \\
\hline
Mean & 126.1 & 56.8 & 196.3 & 294.5 & 0.33 & 0.36 & 0.05 & -0.77 & -2.74 & -7.12 \\
\end{tabular}

\caption{Prediction metrics from 2024-05-10 12:00 to 2025-05-11 12:00 for SWMF, MAGE, and OpenGGCM for all sites where $\Delta\mathbf{B}$ measurements are available.}
\label{table:all_sites_dB_crop}
\note[LW]{Added table to examine impacts of geospace model inputs}
\end{table}
\endgroup

\clearpage

\bibliography{main}

\end{document}